\documentclass[journal,draftclsnofoot,onecolumn,12pt,twoside]{IEEEtranTCOM}
\usepackage{amsmath,amsfonts}
\usepackage{algorithmic}
\usepackage{amsthm}
\usepackage{abraces}
\usepackage{algorithm}
\usepackage{array}
\usepackage[caption=false]{subfig}
\usepackage{textcomp}
\usepackage{stfloats}
\usepackage{url}
\usepackage{verbatim}
\usepackage{graphicx}
\usepackage{cite}
\usepackage{wrapfig}
\usepackage{mathtools}
\usepackage{multirow}
\usepackage{verbatim}
\usepackage{color,soul}
\usepackage{amssymb}
\usepackage{setspace}
\usepackage{cases}

\newtheorem{theorem}{Theorem}[]

\newcommand{\eqcite}[2]{[\citenum{#1}, eq. (#2)]}
\newcommand{\thref}[1]{\textbf{Theorem  \ref{#1}}}

\newcommand{\apref}[1]{A{\small PPENDIX} \ref{#1}}
\newcommand{\figref}[1]{Fig.~\ref{#1}}
\newcommand{\subfigref}[2]{Fig.~\ref{#1}\subref{#2}}
\newcommand{\secref}[1]{Sec.~\ref{#1}}
\newcommand{\tabref}[1]{{Table~\ref{#1}}}

\begin{document}
\raggedbottom
\interfootnotelinepenalty=10000
\title{CSK Realization for MC via Spatially Distributed Multicellular Consortia} 
\author{Rinrada Jadsadaphongphaibool, Dadi Bi, and Yansha Deng
\thanks{The authors are with the Department of Engineering, King’s College London, London WC2R 2LS, U.K. (e-mail:\{rinrada.jadsadaphongphaibool, dadi.bi, yansha.deng\}@kcl.ac.uk). (Corresponding author: Yansha Deng). This paper was presented in part at the Ninth Annual ACM International Conference on Nanoscale Computing and Communication Conference, October 2022 \cite{jad2022BCSK}. This work was supported by Engineering and Physical Sciences Research Council (EPSRC), U.K., under Grant EP/T000937/1.}
}

\markboth{}%
{}


\maketitle
\vspace{-2.5cm}
\begin{abstract}
\vspace{-.5cm}
The design and engineering of molecular communication (MC) components capable of processing chemical concentration signals is the key to unleashing the potential of MC for interdisciplinary applications. By controlling the signaling pathway and molecule exchange between cell devices, synthetic biology provides the MC community with tools and techniques to achieve various signal processing functions. In this paper, we propose a design framework to realize any order concentration shift keying (CSK) systems based on simple and reusable single-input single-output cells. The design framework also exploits the distributed multicellular consortia with spatial segregation, which has advantages in system scalability, low genetic manipulation, and signal orthogonality. We also create a small library of reusable engineered cells and apply them to implement binary CSK (BCSK) and quadruple CSK (QCSK) systems to demonstrate the feasibility of our proposed design framework. Importantly, we establish a mathematical framework to theoretically characterize our proposed distributed multicellular systems. Specially, we divide a system into fundamental building blocks, from which we derive the impulse response of each block and the cascade of the impulse responses leads to the end-to-end response of the system. Simulation results obtained from the agent-based simulator BSim not only validate our CSK design framework but also demonstrate the accuracy of the proposed mathematical analysis.
\end{abstract}
\vspace{-0.7cm}
\begin{IEEEkeywords}
\vspace{-0.6cm}
Concentration shift keying, distributed computation, engineered cells, genetic circuits, molecular communication, spatial segregation, synthetic biology.
\end{IEEEkeywords}
\IEEEpeerreviewmaketitle
\vspace{-0.7cm}
\section{Introduction}
Molecular communication (MC) is a biologically-inspired communication paradigm that exchanges information via chemical signals, which facilitates various applications in the fields of healthcare, environment, industry, etc \cite{farsad2016comprehensive}. Over the past decade, the focus of MC has been on the use of communication engineering tools to model, evaluate, and optimize MC systems, such as molecule propagation modeling\cite{pierobon2010physical,Deng2016MolecularReceiver}, communication capacity\cite{kadloor2012molecular,nakano2012channel}, and transmission and detection schemes\cite{kilinc2013receiver}. Despite those advancements in communication theoretical studies, unified and coherent technologies to design and engineer MC components capable of processing chemical concentration signals are less explored, which hinders the development and implementation of interdisciplinary applications \cite{kuscu2019transmitter,soldner2020survey,bi2021survey}.

Like traditional communication systems, modulation-demodulation is one of the primary key signal processing functions for MC to encode information symbols into physical chemical signals. Among various modulation schemes, concentration shift keying (CSK) is the most popular modulation scheme in the MC literature, since it requires only one type of molecule, has low computation complexity, and can be extended to other modulation strategies \cite{Kuran2021ADiffusion}. In \cite{Dadi2022QCSK}, we have realized several CSK modulation and demodulation functionalities via chemical circuits by designing and regulating a set of chemical reactions in microfluidic systems. In particular, it has shown that different concentration levels can be achieved by biological network motif-inspired chemical reactions and by adjusting the geometry of microfluidic systems \cite{bi2020chemical}. The digital signal processing capability of chemical circuits was also demonstrated to be analogous to digital electronics \cite{bi2021digital}, where any order CSK can be achieved. However, those chemical circuits are described by high-level languages and generic species, which requires an extra effort to find molecules to achieve these circuits in real biological scenarios.

The discipline of synthetic biology provides the engineering community with novel tools and techniques to process chemical concentration signals in cells for MC \cite{khalil2010synthetic,bi2021survey}. Leveraging the activation and repression mechanisms of the gene expression process, engineered cells can imitate logic functions \cite{tamsir2011robust, wang2011engineering}, toggle switches \cite{gardner2000construction}, and a memory device \cite{Urrios2016ADevice}. This motivates several works on engineering MC transceivers in a single cell to perform the M-ary amplitude modulation-demodulation function \cite{unluturk2015biotranceiver} or the single parity-check coding-decoding function \cite{Marcone2018Parity-checkCells}.

Although it is natural to engineer signal processing functions in a single cell from the system point of view, the confined volume of a cell can only accommodate a limited number of signal processing functions. It has shown that the maximum number of repressor-based logic gates per cell cannot exceed eight gates in both \textit{Escherichia coli} (\textit{E. Coli}) \cite{nielsen2016genetic} and yeast\cite{gander2017digital}. A high number of logic gates inside a single cell can lead to a metabolic burden, reduced cell viability, and compromised reusability of the cell \cite{nielsen2016genetic}. This substantially constrains the scalability and complexity of signal processing functions needed for practical applications. Luckily, these issues can be addressed by distributed computation in multicellular consortia, which divides complex signal processing functions into different cells with simple functions and wires them together via signaling molecules \cite{macia2012distributed, Al-Radhawi2020DistributedCircuits}. In this way, the distributed computation can ultimately achieve improved programmability and higher computational complexity through multi-cell cooperation \cite{Macia2016ImplementationConsortia}. 

Motivated by the aforementioned studies, the objective of this paper is to achieve CSK modulation and demodulation functions for MC via distributed multicellular consortia. Although we realized the binary CSK (BCSK) system using engineered cells in \cite{jad2022BCSK}, it is unclear how this BCSK design can be extended to other higher orders of CSK schemes. Inspired by this, we propose a general design framework for CSK modulation and demodulation functions. The contributions of this paper are outlined as follows: \begin{itemize}
    \item We generalize and expand our conference version in \cite{jad2022BCSK} by proposing a general design framework to realize any order of CSK system for MC. Significantly, our design framework is based on single-input single-output cells along with spatial segregation in a microfluidic system, which not only provides signal orthogonality but also allows the simple engineered cells to be reused. Thus, we develop a small library of reusable engineered cells and utilize them to achieve the BCSK and quadruple CSK (QCSK) systems according to the general CSK design framework. Unlike the genetic circuits proposed in \cite{unluturk2015biotranceiver, Marcone2018Parity-checkCells}, the gene expressions and molecules of our proposed system are designed and validated in the existing synthetic biology literature. Hence, our design principle can be used to develop other signal processing functions in MC using synthetic cells.
    \item We establish a novel mathematical framework to characterize our proposed BCSK and QCSK systems. This is accomplished by dividing the system into fundamental building blocks, deriving the impulse response of each building block, and finally deriving the end-to-end expression of the systems. Importantly, our mathematical framework can be applied to analyze other new and more complicated microfluidic-based multicellular circuits.
    \item We present the implementation of our proposed BCSK and QCSK systems in the BSim, which is an agent-based simulator that provides a different set of tools for an efficient simulation of cellular behavior in realistic environmental geometries. The simulation results obtained from BSim validate not only the functionalities of our proposed BCSK and QCSK systems but also our derived mathematical framework. We also evaluate the communication performance of the QCSK system in terms of bit error rate (BER) through BSim simulations. Our result shows that optimizing the detection thresholding value and bit interval can minimize intersymbol interference (ISI).
\end{itemize}

The remainder of this paper is organized as follows. In \secref{sec:DesignFramework}, we propose a general design framework to realize any order of CSK via distributed computation. In \secref{sec:DABQS}, we apply the general CSK design framework to realize the BCSK and QCSK systems and establish a mathematical framework to characterize our proposed systems. Numerical results in \secref{sec:NumericalResults} validate the proposed CSK framework and their theoretical analyses. Finally, we conclude this paper in \secref{sec:Conclusion}.

\vspace{-0.7cm}
\section{Design Framework}\label{sec:DesignFramework}
In this section, we first propose a general design framework to realize any order of CSK using genetic circuits in \secref{sec:DesignPrinciple}. The main idea of the design framework is to implement the CSK function through a small library of reusable engineered cells, where the cells are distributed in different confined spaces and interconnected by signaling molecules to perform signal processing. Then, we present the fundamental processes of molecular signal processing inside an engineered cell in \secref{sec:EGC} and characterize the propagation environment of information molecules in \secref{sec:MCmodel}.  
\vspace{-0.7cm}
\subsection{Design Principle}\label{sec:DesignPrinciple}
\subsubsection{CSK Modulation}
The CSK modulator encodes $m$ bits into $2^m$ different concentrations of a transmitted signal, which is equivalent to the amplitude shift keying (ASK) in conventional communications\cite{kuran2011modulation}. The key challenge of implementing CSK modulation for MC is how to control the output concentration via the $2^m$ different input combinations. We address this by expressing the output concentration of a transmitted signal $C_{\mathrm{Tx}}$ through the base conversion method\footnote{Here, the base conversion in \eqref{eq:modOut} converts binary values to decimal values. For instance, the binary value 11 can be converted to a decimal value $1\times 2^1 + 1\times 2^0=3$.}\cite{kettani2006conversion}, which is \vspace{-0.25cm}\begin{align}
    C_{\mathrm{Tx}} = \sum\limits_{i=0}^{m-1}\varepsilon_{i}S_{i}, \label{eq:modOut}
\end{align} where $S_{i}$ is the $i$th bit of the input $\{ S_{0},S_{1},\cdots,S_{m-1}\}$ and $\varepsilon_{i}$ is the bit weighting of $S_{i}$.

Eq.~\eqref{eq:modOut} can be translated into a biological implementation via $m$ multicellular consortia located in different physically confined spaces\cite{Macia2016ImplementationConsortia}. In particular, the engineered cells in each consortium perform an identity (ID) function that produces a type of wiring molecules only in the presence of the input molecules, i.e., $S_i=1$. The concentration of the wiring molecules of a consortium (i.e., the bit weighting multiplication) can be controlled by adjusting the number of engineered cells, this is because the more cells, the higher the concentration \cite{Urrios2018Plug-and-PlayResponses}. Then, by mixing the wiring molecules released from multicellular consortia, we can implement the summation operator, thus achieving the CSK modulation.

\subsubsection{CSK Demodulation} 
The CSK demodulator decodes $2^{m}$ different concentrations of transmitted signals to $m$ bits. For this purpose, the demodulator is separated into a front-end and a back-end module. The front-end module consists of $2^m-1$ biological thresholding units with different thresholding values to distinguish $2^{m}$ concentration levels into binary signals $B_{j}\; (0\leq j<2^{m}-1)$. In particular, the engineered cells in each thresholding unit produce a type of wiring molecules representing bit-1 only when the input concentration $C_{\mathrm{Tx}}$ exceeds a thresholding value.

\begin{table}[!t]
    \centering
    \caption{CSK Demodulator back-end Truth Table}
    \label{tab:csktt} \vspace{-0.5cm}
    {\small\begin{tabular}{c|ccccc|cccc}
         \hline
         Symbol Index&$B_{2^{m}-2}$&$\cdots$&$B_{2}$&$B_{1}$&$B_{0}$&$Y_{m-1}$&$\cdots$&$Y_{1}$&$Y_{0}$\\\hline
         0& 0&$\cdots$&0&0&0& 0&$\cdots$&0&0\\
         1& 0&$\cdots$&0&0&1& 0&$\cdots$&0&1\\
         2& 0&$\cdots$&0&1&1& 0&$\cdots$&1&0\\
         3& 0&$\cdots$&1&1&1& 0&$\cdots$&1&1\\
         $\vdots$& $\vdots$&$\ddots$&$\vdots$&$\vdots$&$\vdots$& $\vdots$&$\ddots$&$\vdots$&$\vdots$\\
         $2^{m}-1$& 1&$\cdots$&1&1&1& 1&$\cdots$&1&1\\
         \hline
    \end{tabular}}
\end{table}

As the front-end module is only capable of decoding $2^m$ concentration levels to $2{^m}-1$ binary signals, the back-end module is required to further process the signals into $m$ bits. The back-end module extracts demodulated bit $Y_{i}\; (0\leq i < m)$ from $B_{j}$ through logic computation, where the relationship between $Y_{i}$ and $B_{j}$ is presented in \tabref{tab:csktt}. This table describes the output according to all possible input combinations, which can be transformed into Boolean equations in terms of sum-of-product (SOP) expression as \begin{align}
    \begin{cases}
        Y_{0} = (\overline{B_{2^{m}-2}}\cdots\overline{B_{2}}\;\overline{B_{1}}B_{0})+(\overline{B_{2^{m}-2}}\cdots{B_{2}}{B_{1}}B_{0})+\cdots+({B_{2^{m}-2}}\cdots{B_{2}}{B_{1}}B_{0})\\
        Y_{1} = (\overline{B_{2^{m}-2}}\cdots\overline{B_{2}}{B_{1}}B_{0})+(\overline{B_{2^{m}-2}}\cdots{B_{2}}{B_{1}}B_{0})+\cdots+({B_{2^{m}-2}}\cdots{B_{2}}{B_{1}}B_{0})\\
        \;\;\;\;\;\;\;\;\;\;\;\;\;\;\;\;\;\;\;\vdots\\
        Y_{m-1} = (B_{2^{m}-2}\cdots{B_{2}}{B_{1}}B_{0})
    \end{cases}, \label{eq:SOP}
\end{align} where $\overline{B_{j}}$ is the NOT operator of $B_{j}$ and $B_{j}B_{i}$ is the AND operation between $B_{j}$ and $B_{i}$. To simplify the genetic engineering of the cells, we can rewrite \eqref{eq:SOP} into an alternative expression excluding the AND logic based on the inverted logic formulation (ILF) \cite{Macia2016ImplementationConsortia}. According to De Morgan's Law \cite{ciletti2007digital}, ILF is achieved by applying double negations. Thus, eq. \eqref{eq:SOP} can be rewritten as \vspace{-0.3cm}\begin{align}
    Y_{i} = \begin{cases}
    B_{2^{(m-1)}-1},& i = m-1\\
    B_{q}+\sum\limits_{l=1}^{2^{(m-i-1)}-1} (\overline{B_{q - (2l-1)(2^{i})} + \overline{B_{q - 2l(2^{i})}}}),&0\leq i<m-1
    \end{cases}, \label{eq:Yi}
\end{align} where $q = 2^{(m-1)}-1+\sum\limits_{j=0}^{m-i-2}2^{(m-(j+2))}$.
 
The back-end module, described in \eqref{eq:Yi}, can be translated into similar multicellular consortia as CSK modulator. Apart from the ID function and summation operation, engineered cells in the consortium also perform a simple NOT function by inhibiting the production of wiring molecules in the presence of the input molecules. Hereby, we can achieve the back-end module and realize the CSK demodulation by connecting the modules together. 
\vspace{-0.5cm}
\subsection{Fundamental Processes of the Engineered Cells}\label{sec:EGC}
\begin{figure}
    \centering
    \includegraphics[width=0.4\linewidth]{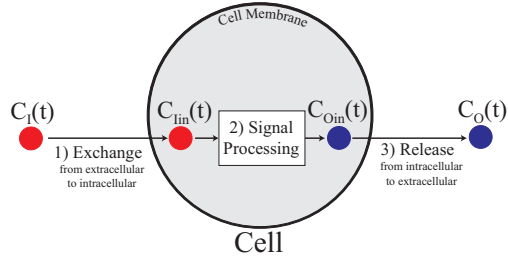}
    \caption{Schematic representation of molecular processing by a cell.}
    \label{fig:GCprocess}
\end{figure}

From the discussion of our proposed design framework, the realization of CSK modulation and demodulation requires three types of engineered cells: ID gate, NOT gate, and thresholding unit. As shown in \figref{fig:GCprocess}, the molecule signal processing by all these three engineered cell types is based on three fundamental processes:  1) the \textit{exchange} of input molecules from the extracellular to the intracellular environment, 2) the \textit{signal processing} of output molecules through regulated gene expressions, and 3) the \textit{release} of the output molecules from the intracellular to the extracellular environment. In the following, we provide the details of these three processes along with their mathematical descriptions.

\subsubsection{Molecule exchange}
When external molecules are absorbed on the cell membrane, the molecules are then transferred from the extracellular to the intracellular environment. The dynamics of molecule exchange can be expressed as \eqcite{fiore2017silico}{7}\begin{align}
    \dfrac{\mathrm{d}C_{\mathrm{Iin}}(t)}{\mathrm{d}t} = \eta C_{\mathrm{I}}(t) - k_{\mathrm{d}}C_{\mathrm{Iin}}(t) \label{eq:CinODE},
\end{align} where $C_{\mathrm{Iin}}(t)$ is the concentration of input molecules inside a cell, $\eta$ is the exchange rate, $C_{\mathrm{I}}(t)$ is the general input molecules and here it represents the molecules in the extracellular environment, and $k_{\mathrm{d}}$ is the degradation rate of the molecules.

\subsubsection{Intracellular signal processing} 
Cells process signals through the gene regulatory network (GRN), which is a set of interacted gene expressions that control a specific cell function. The gene expression mainly consists of two steps: \textit{transcription} and \textit{translation} \cite{Urrios2016ADevice}. First, gene expression begins with transcription, where the RNA polymerase (RNAP) recognizes and binds to the promoter region found at the beginning of a gene to initiate the synthesis of a strand of mRNA. Once mRNA is produced, a ribosome uses tRNA to convert mRNA into protein via translation. The gene expression can be controlled by transcription factors (TFs), which can up-regulate (or down-regulate) the protein production by enhancing (or inhibiting) the binding of RNAP to the promoter site. By designing the GRN, we can realize the ID, NOT, and thresholding signal processing functions.\begin{figure}
    \centering
    \includegraphics[scale = 0.08]{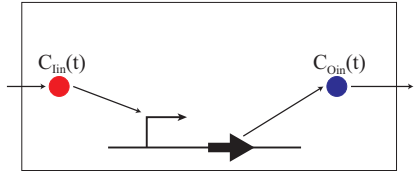}
    \caption{The GRN of the ID gate. The promoter (bent arrow) indicates an area where transcription starts. The protein coding sequence (thick arrow) indicates the region of DNA directly read during translation to produce proteins.}
    \label{fig:GCID}
\end{figure}
\begin{itemize}
    \item \textbf{ID Gate}: As shown in \figref{fig:GCID}, we engineer the ID gate through inducing the gene expression directly by the input molecules (i.e., acting as an activator). Thus, the dynamics of output molecule concentration can be expressed as \eqcite{Ingalls2013MathematicalIntroduction}{7.8}\begin{align}
            \dfrac{\mathrm{d}C_{\mathrm{Oin}}(t)}{\mathrm{d}t} = \beta\dfrac{C_{\mathrm{Iin}}(t)^{n}}{1+[\theta C_{\mathrm{Iin}}(t)]^{n}} - (k_{\mathrm{d}} + \xi )C_{\mathrm{Oin}}(t) \label{eq:CoiODE},
        \end{align} where $C_{\mathrm{Oin}}(t)$ is the concentration of the produced molecules inside a cell, $\beta$ is the protein production rate, $\theta$ is a data fitting parameter, $n$ is the Hill coefficient, and $\xi $ is the rate of the product molecules released from the gate cell.
    \begin{figure}
        \centering
        \begin{minipage}{0.63\linewidth}
            \centering
            \includegraphics[scale = 0.08]{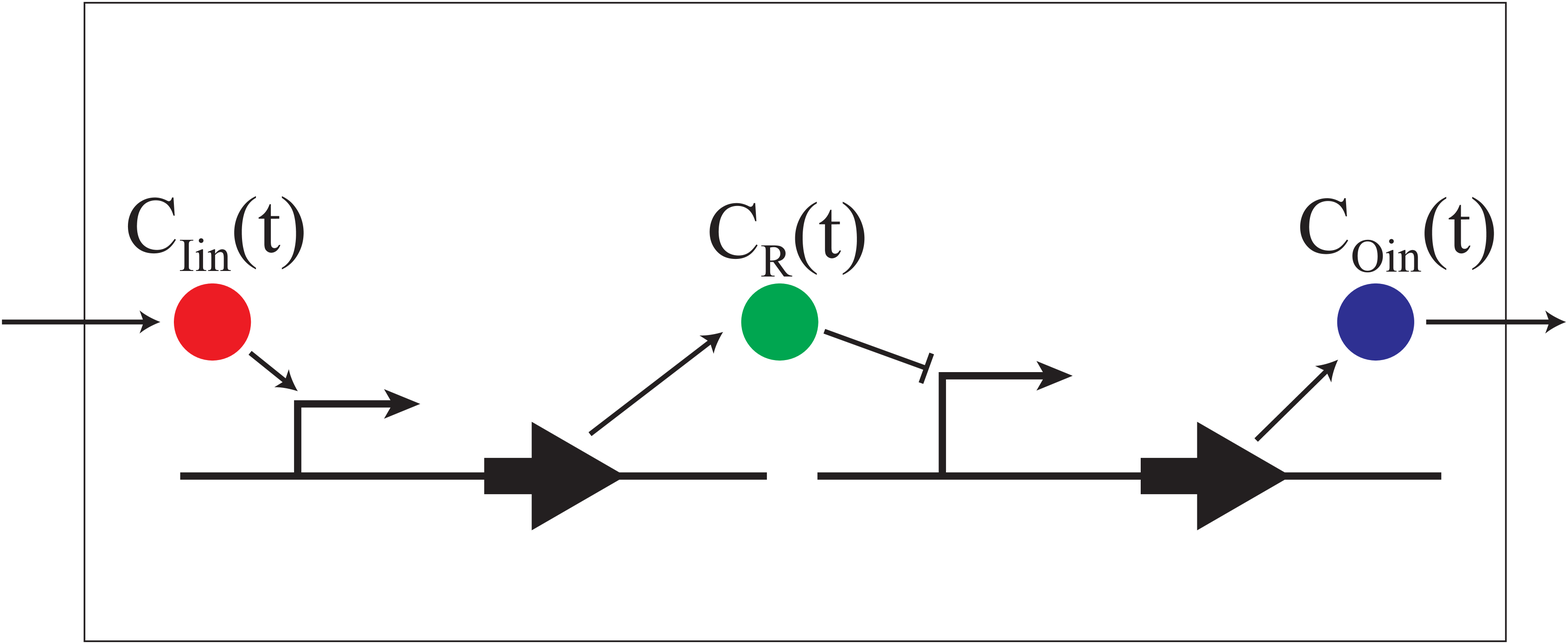}
            \caption{The GRN of the NOT gate. The repression by the binding of TFs (indicated with a flat head arrow) prevents the expression of downstream proteins.}
            \label{fig:GCNOT}
        \end{minipage}\hspace{0.5cm}
        \begin{minipage}{0.27\linewidth}
            \centering
            \includegraphics[scale = 0.08]{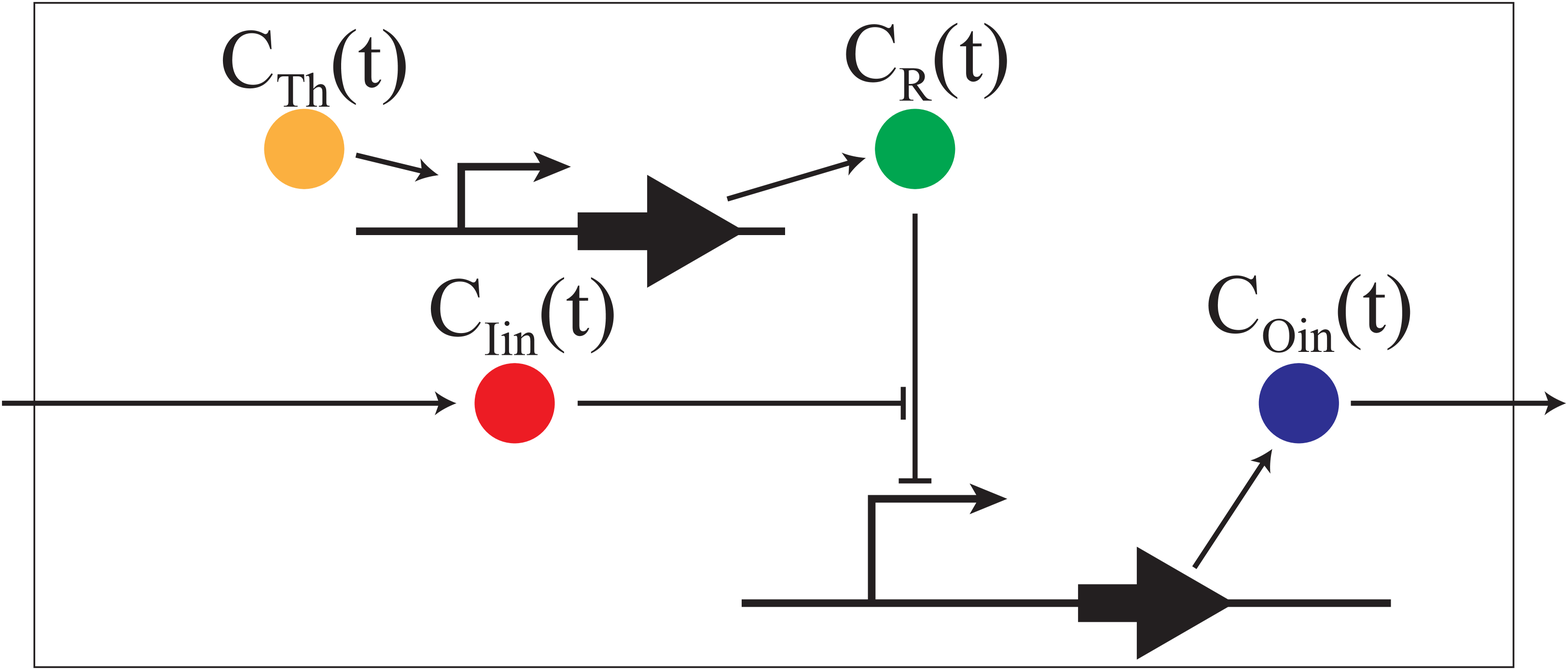}
            \caption{The GRN of the thresholding unit.}
            \label{fig:GCTH}
        \end{minipage}
    \end{figure}
    \item \textbf{NOT Gate}: To implement the NOT function, we borrow an optimal configured and experimentally validated GRN of the engineered yeast cells from \cite{Urrios2016ADevice}. As shown in Fig. 3, the presence of the input molecules activates the production of a repressor protein, which can be modeled by \vspace{-0.3cm} \begin{align}
        \dfrac{\mathrm{d}C_{\mathrm{R}}(t)}{\mathrm{d}t} = \beta\dfrac{C_{\mathrm{Iin}}(t)^{n}}{1+[\theta C_{\mathrm{Iin}}(t)]^{n}} - k_{\mathrm{d}}C_{\mathrm{R}}(t), \label{eq:dCR3}
    \end{align} where $C_{\mathrm{R}}(t)$ is the concentration of the repressor protein. Then, the produced repressor protein binds to the downstream gene and precludes the expression of the output molecules, which can be expressed as \eqcite{Ingalls2013MathematicalIntroduction}{7.10}\begin{align}
        \dfrac{\mathrm{d}C_{\mathrm{Oin}}(t)}{\mathrm{d}t} = \beta\dfrac{1}{1+[\theta C_{\mathrm{R}}(t)]^{n}} - (k_{\mathrm{d}}+\xi )C_{\mathrm{Oin}}(t). \label{eq:COin3}
    \end{align}
    \item \textbf{Thresholding Unit}: As shown in \figref{fig:GCTH}, different from the GRN of the NOT gate, the input molecules react with the repressor protein induced by the threshold molecules, thus controlling the amount of the output molecules. The reaction between the input molecules and the repressor can be expressed as  \begin{align}
        \mathrm{I}_{\mathrm{in}} + \mathrm{R} \xrightarrow[]{k_{\mathrm{f}}} \phi, \label{eq:chemreact}
    \end{align} where $k_{\mathrm{f}}$ is the forward reaction rate constant and $\phi$ is a type of molecule irrelevant for signal processing. As such, the dynamics of input molecule concentration in \eqref{eq:CinODE} can be rewritten as  \begin{align}
         \dfrac{\mathrm{d}C_{\mathrm{Iin}}(t)}{\mathrm{d}t} = \eta C_{\mathrm{I}}(t) - k_{\mathrm{f}}C_{\mathrm{Iin}}(t)C_{\mathrm{R}}(t) - k_{\mathrm{d}}C_{\mathrm{Iin}}(t), \label{eq:dCIin4}
     \end{align} and the dynamics of repressor protein concentration can be expressed as \eqcite{Jamali2019Channel}{22a} \vspace{-0.3cm}\begin{align}
        \dfrac{\mathrm{d}C_{\mathrm{R}}(t)}{\mathrm{d}t} = f_{\mathrm{R}}(t) - k_{\mathrm{f}}C_{\mathrm{Iin}}(t)C_{\mathrm{R}}(t) - k_{\mathrm{d}}C_{\mathrm{R}}(t), \label{eq:dCR4}
    \end{align} where $f_{\mathrm{R}}(t) = \beta\frac{C_{\mathrm{Th}}(t)^{n}}{1+[\theta C_{\mathrm{Th}}(t)]^{n}}$ is the rate of repressor production and $C_{\mathrm{Th}}(t)$ is the concentration of the threshold molecules. The thresholding value of the cell is the concentration of the repressor molecules in equilibrium without the consumption by the input molecules, i.e., $f_{\mathrm{R}}(t) -k_{\mathrm{d}}C_{\mathrm{R}}(t) = 0$ in \eqref{eq:dCR4}. Hence, the thresholding value can be tuned by controlling the production of repressor via $C_{\mathrm{Th}}(t)$. Combining \eqref{eq:COin3} with \eqref{eq:dCR4}, we can express the signal processing of a thresholding unit.
\end{itemize}
\subsubsection{Molecule release}
With the production of the output molecules, they are further released from the intracellular to the extracellular. The dynamics of molecule release can be expressed as \eqcite{fiore2017silico}{7} \begin{align}
    \dfrac{\mathrm{d}C_{\mathrm{O}}(t)}{\mathrm{d}t} = \xi C_{\mathrm{Oin}}(t)\label{eq:CoutODE},
\end{align} where $C_{\mathrm{O}}(t)$ is the concentration of the released molecules from cells.
\vspace{-0.5cm}
\subsection{Propagation of Signaling Molecules}\label{sec:MCmodel}
Under the multicellular consortium setting, it is essential to establish communication between different consortia to achieve community level functions (e.g., CSK modulation and demodulation). To provide wiring (or signaling) molecules with guided transmission and increased propagation speed, we choose to incubate engineered cells in a microfluidic system. The integration of engineered cells with a microfluidic system not only provides an opportunity to realize spatial segregation by isolating different cell types at different locations \cite{Macia2016ImplementationConsortia, Dadi2022QCSK}, but also avoids the cross-talk between cells by controlling the fluid direction. 

\begin{figure}
    \centering
    \includegraphics[width = 0.5\linewidth]{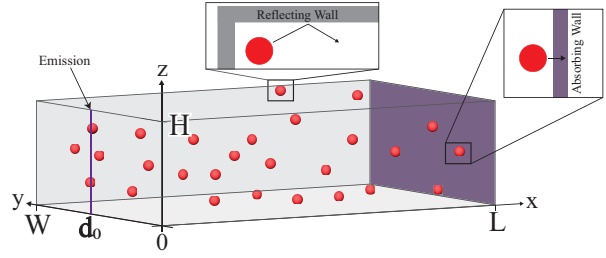}
    \caption{The emission, propagation, and reception of signaling molecules.}
    \label{fig:MC}
\end{figure}

For a 3D microfluidic channel, as shown in Fig. 5, the molecular concentration can be described by the advection-diffusion-reaction as \eqcite{Jamali2019Channel}{29}
\begin{align}
    \dfrac{\partial C(\mathbf{d},t)}{\partial t} =& D\nabla^{2}C(\mathbf{d},t) - \nabla\cdot[\mathbf{u}(\mathbf{d})C(\mathbf{d},t)]- k_{\mathrm{d}}C(\mathbf{d},t) \label{eq:ard},
\end{align} where $\mathbf{d} = [x, y, z]$ is a vector defining the position in space, $D$ is the diffusion constant, $\nabla^{2}$ is the Laplace operator, $\nabla$ is the Nabla operator, and $\mathbf{u}(\mathbf{d})$ is the local velocity. The flow inside the microfluidic channel is assumed to be laminar, which is a valid assumption in microfluidic settings \cite{kumar2010microfluidic}. For the tractability of the analysis, we further assume a uniform flow with constant velocity $u$ along the $y$- and $z$-axis \cite{Mosayebi2019EarlyNanosensors}. As a result, eq. \eqref{eq:ard} can be reduced to
\begin{align}
    \dfrac{\partial C(\mathbf{d},t)}{\partial t} = D\nabla^{2}C(\mathbf{d},t) - u\dfrac{\partial C(\mathbf{d},t)}{\partial x} - k_{\mathrm{d}}C(\mathbf{d},t) \label{eq:ard2}.
\end{align}

When molecules are uniformly released over $\mathbf{d}_{0} = [0, y_{0}, z]$ into the microfluidic channel at time $t=0$, the initial condition can be defined as \begin{align}
    C(\mathbf{d}_{0},t)\big|_{t=0} = \delta(x)\delta(y-y_{0}) \label{eq:IC},
\end{align} where $\delta(\cdot)$ is the Dirac delta function. 

As the molecules propagate along the $x$-axis, the boundaries of the microfluidic channel are assumed to be reflective, and we can define their boundary conditions as \eqcite{Jamali2019Channel}{8}\begin{align}
    \dfrac{\partial C(\mathbf{d},t)}{\partial y}\bigg|_{y=0, W} = 0 \label{eq:yBC},
\end{align}\begin{align}
    \dfrac{\partial C(\mathbf{d},t)}{\partial z}\bigg|_{z=0, H} = 0 \label{eq:zBC},
\end{align} where $W$ and $H$ are the width and height of the channel, respectively.

The molecules propagated inside the microfluidic channel can be received by a column of well-aligned engineered cells located at $L + R$, where $L$ is the distance between the releasing point and the cell and $R$ is the radius of the cell. In this paper, the used signaling molecules (provided in \secref{sec:DABQS}) are too large to cross the cell membrane, and their reception relies on the receptors distributed on the surface of the cell membrane \cite{bi2021survey, Deng2016MolecularReceiver}. Therefore, each engineered cell can be modeled as an active receiver, and we assume the well-aligned cells create an absorbing wall along the $y$-$z$ plane. For simplicity, we assume that the molecules will be absorbed at the left side of the cells (e.g., $x=L$), and the molecules produced by GRN will be released at the right side of the cells (e.g., $x=L + 2R$). Hence, another boundary condition can be defined as \eqcite{Deng2016MolecularReceiver}{5} \vspace{-0.3cm}\begin{align}
    D\dfrac{\partial C(\mathbf{d},t)}{\partial x}\bigg|_{x \xrightarrow[]{} L} - k_{\mathrm{a}}C(\mathbf{d},t)\bigg|_{x \xrightarrow[]{} L} = 0\label{eq:AbWallBC},
\end{align} where $k_{\mathrm{a}}$ is the absorption rate of one cell.

\vspace{-0.5cm}
\section{Design and Analysis of BCSK and QCSK Systems}\label{sec:DABQS}
\begin{figure}
    \centering
    \includegraphics[width = 0.5\linewidth]{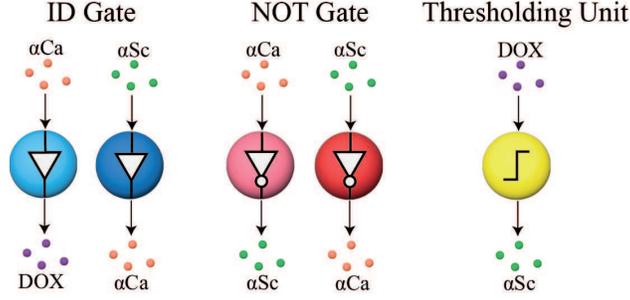}
    \caption{The library of engineered cells used to achieve BCSK and QCSK.}
    \label{fig:celllib}
\end{figure}
In this section, we apply the general CSK design framework to realize the BCSK and QCSK systems to demonstrate the feasibility of the design principle. In order to implement the BCSK and QCSK systems according to the design principle presented in \secref{sec:DesignPrinciple}, we develop a library of engineered yeast cells as shown in \figref{fig:celllib}. This library consists of five types of cells that achieve three signal processing functions by responding and producing three signaling molecules: mating $\alpha$-factor from \textit{C.albicans} ($\alpha$Ca), mating $\alpha$-factor from \textit{Saccharomyces cerevisiae} ($\alpha$Sc), and doxycycline (DOX). In particular, the inputs and outputs of the engineered cells are:  \begin{itemize}
    \item \textbf{ID Gate:} responds to $\alpha$Ca and $\alpha$Sc and produces DOX and $\alpha$Ca, respectively;
    \item \textbf{NOT Gate:} responds to $\alpha$Ca and $\alpha$Sc to express the LacI repressor (see \figref{fig:GCNOT}) that prevents the expression of $\alpha$Sc and a$\alpha$Ca, respectively;
    \item \textbf{Thresholding Unit:} responds to DOX, which reacts with TetR repressor (see \figref{fig:GCTH}) and inhibits the repression of $\alpha$Sc.
\end{itemize}
In the following, we first present how to use the cell library to achieve the BSCK and QCSK designs in \secref{sec:DPS}. We then analyze the building blocks of the BCSK and QCSK systems in \secref{sec:CA} and apply them to examine the output responses of the systems in \secref{sec:APS}.
\vspace{-0.5cm}
\subsection{Design of BCSK and QCSK Systems}\label{sec:DPS}
\subsubsection{BCSK Design}
\begin{figure}
    \centering
    \includegraphics[width=0.8\linewidth]{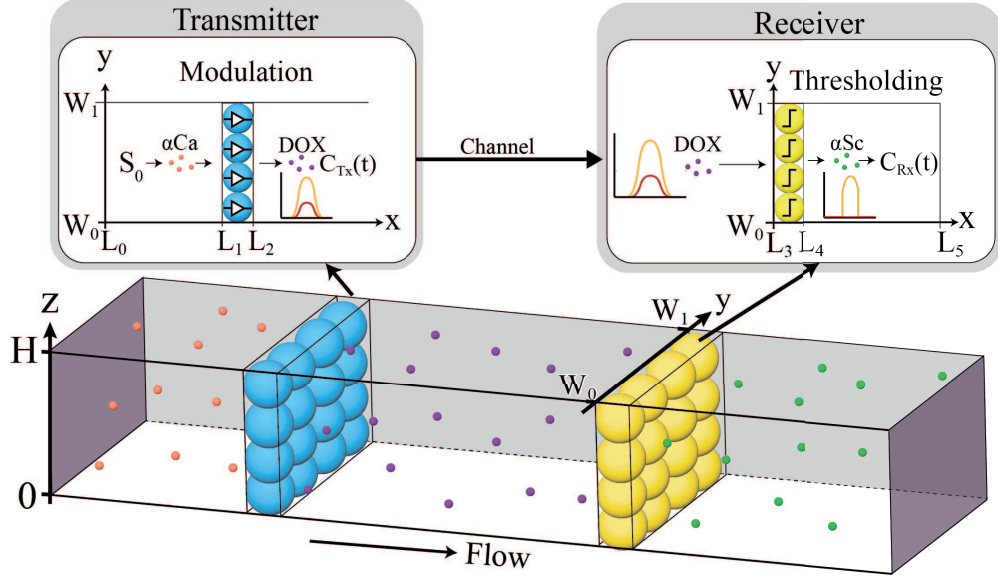}
    \caption{The implementation of the BCSK system.}
    \label{fig:BCSK}
\end{figure}
The BCSK system encodes one bit of information into two different concentration levels; the BCSK is also known as on-off keying (OOK) when zero molecules are released to send a bit-0. We present our proposed OOK design in \figref{fig:BCSK}. The OOK transmitter employs the ID cell ($\alpha$Ca-DOX) to respond to the input bit $S_0$, represented by molecules $\alpha$Ca, with a concentration $C_{S_0}(t)$. The reception of $\alpha$Ca molecules by the ID cells can lead to the expression and release of DOX into the microfluidic channel, generating the modulated signals denoted by $C_{\mathrm{Tx}}(t)$. Due to the ID logic of the engineered cells, no molecules will be released for bit-0 transmission, thus achieving the OOK modulation.

Once the DOX molecules arrive at the receiver, they can activate the thresholding cell (DOX-$\alpha$Sc) to produce $\alpha$Sc molecules and a binary signal $B_{0}$. For $m = 1$, eq. \eqref{eq:Yi} can be expressed as $Y_{0} = B_{0}$; therefore, the thresholding unit is sufficient to achieve OOK demodulation. The $\alpha$Sc molecules then propagate to the detection wall, which generates the final output of Rx, denoted by $C_{\mathrm{Rx}}(t)$. In practice, the concentration level can be visualized by fluorescent proteins or measured by electrochemical sensors experimentally.

\subsubsection{QCSK Design}
\begin{figure}
    \centering
    \includegraphics[width=0.9\linewidth]{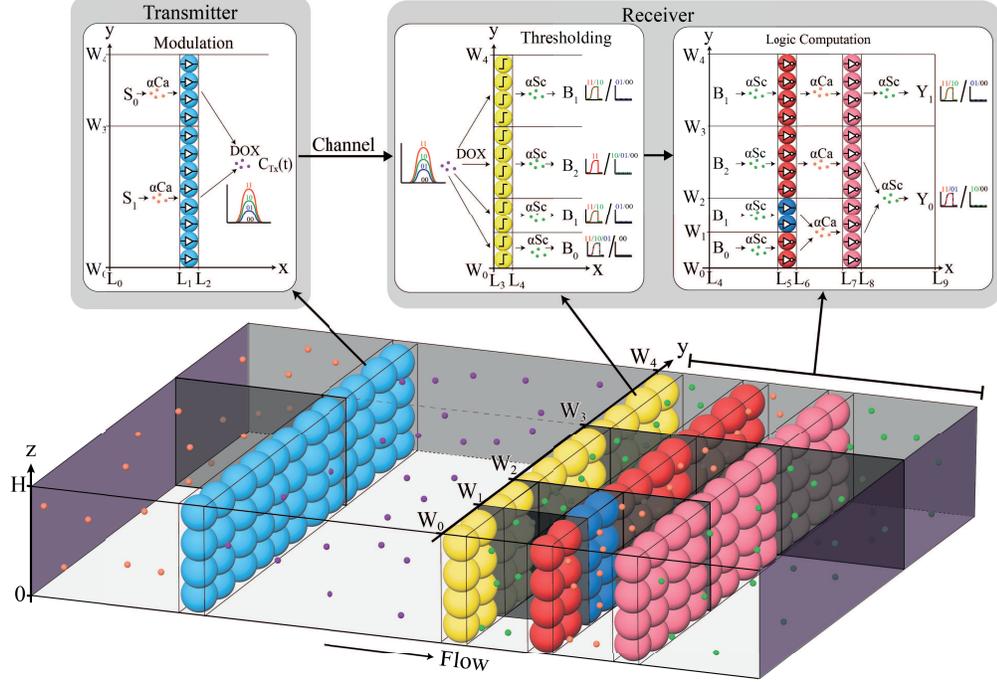}
    \caption{The implementation of the QCSK system.}
    \label{fig:QCSK}
\end{figure}
The QCSK system encodes two bits of information into four concentration levels. \figref{fig:QCSK} presents our proposed QCSK design. For $m=2$, eq. \eqref{eq:modOut} can be rewritten as \begin{align}
    C_{\mathrm{Tx}} = \varepsilon_{0}S_{0} + \varepsilon_{1}S_{1}.\label{eq:Ctxm2}
\end{align} The QCSK modulator encodes input bits, $S_{1}$ and $S_{0}$, by employing two populations of ID cells ($\alpha$Ca-DOX). The number of cells in the population responding to $S_1$ is twice of that in the other population, i.e., $\varepsilon_{1}=2 \varepsilon_{0}$ in \eqref{eq:Ctxm2}. By doing so, bit $S_1$ leads to a higher concentration output than bit $S_0$ and the combination of $S_1$ and $S_0$ produces four concentration levels, thus achieving the QCSK modulation.

At the receiver side, DOX molecules activate three populations of thresholding cells (DOX-$\alpha$Sc) to distinguish the four concentration levels into $B_{0}$, $B_{1}$, and $B_{2}$. For $m=2$, eq. \eqref{eq:Yi} can be reduced to \vspace{-0.7cm}\begin{align}
    Y_{1} = B_{1} = \overline{\overline{B_{1}}}, \label{eq:Y1boolean}
\end{align}\vspace{-1cm}\begin{align}
    Y_{0} = B_{2} + (\overline{B_{1} + \overline{B_{0}}}) =  \overline{\overline{B_{2}}} + (\overline{B_{1} + \overline{B_{0}}}). \label{eq:Y0boolean}
\end{align} 
Although \eqref{eq:Y1boolean} reveals that the output of the thresholding unit $B_{1}$ can be directly interpreted as $Y_{1}$, we further apply a double NOT operation to $B_{1}$ to ensure that $Y_{1}$ has the same concentration level as $Y_{0}$. Specifically, the first NOT cells ($\alpha$Sc-$\alpha$Ca) take the thresholding output signal $B_{1}$, which inhibits the production of $\alpha$Ca molecules. The output of the first NOT cells propagates to the second population of NOT cells ($\alpha$Ca-$\alpha$Sc), where $\alpha$Ca inhibits the production of $\alpha$Sc molecules and output $Y_{1}$. For \eqref{eq:Y0boolean}, three thresholding signals are processed and combined to output $Y_{0}$. The thresholding output signal $B_{2}$ experiences the same double NOT operation as the $B_{1}$ for $Y_{1}$. Simultaneously, the NOT cells ($\alpha$Sc-$\alpha$Ca) and ID cells ($\alpha$Sc-$\alpha$Ca) that respectively receive $B_{0}$ and $B_{1}$ signals from the thresholding cells, are employed to output $\alpha$Ca into the same propagation channel; the mixture of $\alpha$Ca molecules gives $B_{1} + \overline{B_{0}}$. Then, the $\alpha$Ca molecules are processed by NOT cells ($\alpha$Ca-$\alpha$Sc) and the output is mixed with $\overline{\overline{B_{2}}}$ to obtain $Y_{0}$. 
\vspace{-0.5cm}
\subsection{Circuit Analysis}\label{sec:CA}
To ease the analysis of the CSK design, we divide the end-to-end CSK system into four building blocks: ID block, NOT block, thresholding block, and propagation block. For the first three blocks, each block can contain one or more corresponding engineered cells. Recall that the engineered cell consists of three processes represented by ordinary differential equations (ODEs) in \secref{sec:EGC}, therefore each cell can be described by a set of ODEs. As all three blocks experience the same molecule exchange, which is independent of any other processes, we first derive the impulse response of the molecule exchange in \eqref{eq:CinODE} as \begin{align}
    h_{1}(t) = \eta e^{-k_{\mathrm{d}}t}\label{eq:solutionCin}.
\end{align} By solving the other processes of the engineered cell along with \eqref{eq:solutionCin}, we obtain the output of the first three blocks in the following. We then derive the output of the propagation block, which is modeled by a partial differential equation (PDE) and a set of constraints in \secref{sec:MCmodel}.

\subsubsection{ID Block}
The ID block is governed by the ODEs in \eqref{eq:CinODE}, \eqref{eq:CoiODE}, and \eqref{eq:CoutODE}. Relying on \eqref{eq:solutionCin} and the Laplace transform, we solve this set of ODEs and obtain the accumulated concentration of the output molecules emitted by ID block as \begin{align}
    C_{\mathrm{O}}^{\mathrm{ID}}(t) = \dfrac{\beta  \xi [1-e^{-(k_{\mathrm{d}} + \xi )t}]}{\theta^{n}(k_{\mathrm{d}} + \xi )} * \Bigg\{1 - \dfrac{1}{1+[\theta C_{\mathrm{I}}(t) *h_{1}(t)]^{n}}\Bigg\}\label{eq:solutionCout},
\end{align} where $*$ is the convolution operator. We also define operator $\mathcal{I}[\cdot]$ to characterize the expected net change of the output concentration released on the surface of the cells during the interval $[t,t+t_{s}]$ as\vspace{-0.5cm}\begin{align}
    \mathcal{I}[C_{\mathrm{I}}(t),I,O] =& C_{\mathrm{O}}^{\mathrm{ID}}(t+t_{s}) - C_{\mathrm{O}}^{\mathrm{ID}}(t),
\end{align} where $I$ is the input molecule type, $O$ is the output molecule type, and $t_s$ is the sampling time. As different molecule types have different production rates, degradation rates, Hill function coefficients, and parameter fitting data, the variables $I$ and $O$ determine the values of $\beta$, $\theta$, $n$, and $k_{d}$.

\subsubsection{NOT Block}
The NOT block is defined by the ODEs in \eqref{eq:CinODE}, \eqref{eq:dCR3}, \eqref{eq:COin3}, and \eqref{eq:CoutODE}. By solving \eqref{eq:dCR3} through Laplace transform, we can derive the concentration of the repressor molecules as \begin{align}
    C_{\mathrm{R}}^{\mathrm{NOT}}(t) =  \dfrac{\beta e^{-k_{\mathrm{d}}t}}{\theta^{n}} * \bigg\{1- \dfrac{1}{1+[\theta C_{\mathrm{I}}(t)*h_{1}(t)]^{n}}\bigg\}. \label{eq:CR3}
\end{align} We then perform Laplace transform on \eqref{eq:COin3} and \eqref{eq:CoutODE} and arrive at the accumulated concentration of the output molecules emitted by NOT block as \begin{align}
    C_{\mathrm{O}}^{\mathrm{NOT}}(t) =\dfrac{\xi\beta[1-e^{-(k_{\mathrm{d}}+\xi )t}]}{k_{\mathrm{d}}+\xi} * \dfrac{1}{1+[\theta C_{\mathrm{R}}^{\mathrm{NOT}}(t)]^{n}}. \label{eq:CO3a}
\end{align} We define operator $\mathcal{N}[\cdot]$ to characterize the net change of the output concentration and this can be expressed based on \eqref{eq:CO3a} as \begin{align}
    \mathcal{N}[&C_{\mathrm{I}}(t),I,O] = C_{\mathrm{O}}^{\mathrm{NOT}}(t+t_{s}) - C_{\mathrm{O}}^{\mathrm{NOT}}(t).
\end{align} 

\subsubsection{Thresholding Block}
The thresholding block is described by the ODEs in \eqref{eq:CinODE}, \eqref{eq:COin3}, \eqref{eq:dCR4}, and \eqref{eq:CoutODE}. Due to the non-linearity of \eqref{eq:dCR4} caused by the bimolecular reaction, we numerically solve the thresholding block by discretizing the processes into many time intervals with step $t_{s}$, such that $C_{\mathrm{R}}^{\mathrm{Th}}[m] = C_{\mathrm{R}}^{\mathrm{Th}}(mt_{s})$, in the following theorem. \vspace{-0.5cm}\begin{theorem}\label{th:CR4}
    The concentration of the repressor at time $mt_{s}$ can be approximated as\begin{align}
        C_{\mathrm{R}}^{\mathrm{Th}}[m] \approx \begin{cases}
        C_{\mathrm{R_0}}, & C_{\mathrm{R_0}} = 0 \;\mathrm{ or }\; C_{\mathrm{I_0}} = 0\\
        \dfrac{C_{\mathrm{R_0}}(C_{\mathrm{R_0}} - C_{\mathrm{I_0}})}{C_{\mathrm{R_0}}-C_{\mathrm{I_0}}\exp[-k_{\mathrm{f}}t_{s}(C_{\mathrm{R_0}}-C_{\mathrm{I_0}})]},&C_{\mathrm{R_0}}\neq C_{\mathrm{I_0}}\\
        \dfrac{C_{\mathrm{R_0}}}{1+k_{\mathrm{f}}t_{s}C_{\mathrm{R_0}}},&C_{\mathrm{R_0}} = C_{\mathrm{I_0}}
        \end{cases},\label{eq:CR4}
    \end{align} where $C_{\mathrm{R_0}} = \{t_{s}f_{\mathrm{R}}[m-1] + C_{\mathrm{R}}^{\mathrm{Th}}[m-1]\}e^{-k_{\mathrm{d}}t_{s}}$ and $C_{\mathrm{I_0}} = \{t_{s}\eta C_{\mathrm{I}}[m-1] + C_{\mathrm{Iin}}^{\mathrm{Th}}[m-1]\}e^{-k_{\mathrm{d}}t_{s}}$ are the concentrations of the repressor and input molecules at time $(m-1)t_{s}$, respectively. 
\end{theorem}
\vspace{-0.5cm}
\begin{proof}
    See \apref{ap:proofBB4}.
\end{proof}
The concentration of the input molecules is described by the ODE in \eqref{eq:dCIin4}, which has a similar form as \eqref{eq:dCR4}; therefore, according to \thref{th:CR4}, the input concentration at $mt_{s}$ can be derived as \begin{align}
    C_{\mathrm{Iin}}^{\mathrm{Th}}[m] \approx \begin{cases}
    C_{\mathrm{I_0}}, & C_{\mathrm{I_0}} = 0 \;\mathrm{ or }\; C_{\mathrm{R_0}} = 0\\
    \dfrac{C_{\mathrm{I_0}}(C_{\mathrm{I_0}} - C_{\mathrm{R_0}})}{C_{\mathrm{I_0}}-C_{\mathrm{R_0}}\exp[-k_{\mathrm{f}}t_{s}(C_{\mathrm{I_0}}-C_{\mathrm{R_0}})]},&C_{\mathrm{I_0}}\neq C_{\mathrm{R_0}}\\
    \dfrac{C_{\mathrm{I_0}}}{1+k_{\mathrm{f}}t_{s}C_{\mathrm{I_0}}},&C_{\mathrm{I_0}} = C_{\mathrm{R_0}}
    \end{cases}.\label{eq:CIin4}
\end{align} Since the output production and the release processes for the thresholding block are the same as the NOT block, the output concentration emitted by the thresholding block can be expressed as \begin{align}
    C_{\mathrm{O}}^{\mathrm{Th}}(t) =\dfrac{\xi\beta[1-e^{-(k_{\mathrm{d}}+\xi )t}]}{k_{\mathrm{d}}+\xi} * \dfrac{1}{1+[\theta C_{\mathrm{R}}^{\mathrm{Th}}(t)]^{n}}. \label{eq:CO3th}
\end{align} We then characterize the net change of the output concentration of the thresholding block as an operator $\mathcal{T}[\cdot]$, which can be expressed as\begin{align}
    \mathcal{T}[&C_{\mathrm{I}}(t),C_{\mathrm{Th}}(t),I,O] = C_{\mathrm{O}}^{\mathrm{Th}}[t+t_{s}] - C_{\mathrm{O}}^{\mathrm{Th}}(t). \label{eq:T}
\end{align}

\subsubsection{Propagation Block}
The propagation of signaling molecules is governed by the PDE in \eqref{eq:ard2}, with the initial condition in \eqref{eq:IC} and the boundary conditions in \eqref{eq:yBC}-\eqref{eq:AbWallBC}. For the propagation block, we are interested in the number of absorbed molecules after propagating through the channel, which can be derived in the following theorem.
\vspace{-0.5cm}
\begin{theorem}\label{th:prop} After a propagation of a distance $L$, the concentration of the molecules absorbed by a surface $\mathcal{S}_{\mathrm{a}} = [y_{1} \leq y \leq y_{2}, 0\leq z\leq H]$ during $[t, t+t_{s}]$ is derived as \begin{align}
    \mathcal{P}[C_{\mathrm{I}}(t),L,\mathcal{S}_{\mathrm{e}},\mathcal{S}_{\mathrm{a}}] = \dfrac{k_{\mathrm{a}}t_{s}}{H(y_{2}-y_{1})} \big[C_{\mathrm{I}}(t) * h_{2}(t|L, \mathcal{S}_{\mathrm{e}}, \mathcal{S}_{\mathrm{a}}) \big], \label{eq:gateOut}
\end{align} where $C_{\mathrm{I}}(t)$ is the input concentration of the propagation block, $\mathcal{S}_{\mathrm{e}}= [y_{0_{1}} \leq y \leq y_{0_{2}}, 0\leq z\leq H]$ is the emission area along the $y$-$z$ plane, and $h_{2}(t|L, \mathcal{S}_{\mathrm{e}}, \mathcal{S}_{\mathrm{a}})$ is derived as 
   \begin{align}
    h_{2}(t| L,\mathcal{S}_{\mathrm{e}}, \mathcal{S}_{\mathrm{a}}) &= \dfrac{2H}{W}\sum\limits_{l=1}^{\infty} \dfrac{(\lambda^{2}_{l} + G_{1}^{2})\cos(\lambda_{l}L)}{L(\lambda^{2}_{l} + G_{1}^{2})+G_{1}}e^{\frac{u(2L-ut)}{4D}-\lambda_{l}^{2}Dt-k_\mathrm{d}t} \bigg[ (y_{0_{2}} - y_{0_{1}})(y_{2} - y_{1}) \nonumber\\&+ 2\sum\limits_{i=1}^{\infty} [\sin(\gamma_{i}y_{0_{2}})-\sin(\gamma_{i}y_{0_{1}})][\sin(\gamma_{i}y_{2})-\sin(\gamma_{i}y_{1})]e^{-\gamma_{i}^{2}Dt}\bigg], \label{eq:h1boundary}
\end{align} 
where $\lambda_{l}$ is the root of a transcendental equation $\lambda_{l}\tan(\lambda_{l} L) = G_{1}$, $\gamma_{i} = \frac{i\pi}{W}$, $\cos(\cdot)$ is a cosine function, and $G_{1} = \frac{k_{\mathrm{a}}}{D}$.
\end{theorem}
\vspace{-0.5cm}
\begin{proof}
See \apref{ap:proofprop}.
\end{proof}
\vspace{-0.5cm}
\subsection{Analysis of BCSK and QCSK Systems}\label{sec:APS}
Relying on the analyses of the building blocks, we analyze our proposed BCSK and QCSK systems in the following. Let $\mathcal{S}_{i,j}$ denotes a surface on the $y$-$z$ plane with $W_{i} \leq y \leq W_{j}$ and $0 \leq z\leq H$.
\subsubsection{BCSK Analysis}\label{sec:BCSK}
As shown in \figref{fig:BCSK}, the input molecules $\alpha$Ca propagate to a population of ID cells at the transmitter side, and the net concentration of the released molecules by ID cells ($\alpha$Ca-DOX) at $x = L_{2}$ during $[t,t+t_{s}]$ can be expressed as  \begin{align}
    C_{\mathrm{Tx}}(t,t+t_{s}) = \mathcal{I}\Big[\mathcal{P}[C_{\mathrm{S_0}}(t),L_{1}-L_{0},\mathcal{S}_{0,1},\mathcal{S}_{0,1}], \alpha\mathrm{Ca}, \mathrm{DOX} \Big]. \label{eq:BCSKTx}
\end{align}The released DOX molecules propagate inside the channel to a population of thresholding cells, and the net concentration of the molecules absorbed by the thresholding cells at $x = L_{3}$ is derived as \vspace{-0.5cm} \begin{align}
    C_{\mathrm{Ch}}(t,t+t_{s}) = \mathcal{P}[C_{\mathrm{Tx}}(t,t+t_{s}), L_{3}-L_{2},\mathcal{S}_{0,1},\mathcal{S}_{0,1}].
\end{align} The thresholding cells (DOX-$\alpha$Sc) perform BCSK demodulation, where the net concentration of the molecules detected by the detection wall at $x=L_{5}$ can be derived as \begin{align}
    &C_{\mathrm{Rx}}(t,t+t_{s}) = \mathcal{P}\Big[\mathcal{T}[C_{\mathrm{Ch}}(t,t+t_{s}),C_{\mathrm{Th}}(t),\mathrm{DOX},\alpha\mathrm{Sc}],L_{5}-L_{4}, \mathcal{S}_{0,1},\mathcal{S}_{0,1}\Big]. \label{eq:BCSKRx}
\end{align}

\subsubsection{QCSK Analysis}\label{sec:QCSK}
For the QCSK transmitter in \figref{fig:QCSK}, the input bits $S_{1}$ and $S_{0}$ control the total concentration of DOX molecules generated at the transmitter. The net concentration of DOX molecules released by ID cells ($\alpha$Ca-DOX) at $x = L_{2}$ during $[t,t+t_{s}]$ with respect to input bit $S_{1}$ and $S_{0}$ can be expressed as\begin{align}
    C_{\mathrm{Tx}}^{S_1}(t,t+t_{s}) = \mathcal{I}\Big[\mathcal{P}[C_{\mathrm{S_1}}(t),L_{1}-L_{0},\mathcal{S}_{3,4},\mathcal{S}_{3,4}], \alpha\mathrm{Ca}, \mathrm{DOX} \Big],
\end{align}and \begin{align}
    C_{\mathrm{Tx}}^{S_0}(t,t+t_{s}) = \mathcal{I}\Big[\mathcal{P}[C_{\mathrm{S_0}}(t),L_{1}-L_{0},\mathcal{S}_{0,3},\mathcal{S}_{0,3}], \alpha\mathrm{Ca}, \mathrm{DOX} \Big].
\end{align} Therefore, the total net concentration of DOX molecules generated by the QCSK transmitter can be expressed as \vspace{-0.5cm}\begin{align}
    C_{\mathrm{Tx}}(t,t+t_{s}) =& C_{\mathrm{Tx}}^{S_1}(t,t+t_{s})+C_{\mathrm{Tx}}^{S_0}(t,t+t_{s}).\label{eq:QCSKTx}
\end{align} 

Then, the molecules propagate along a channel to populations of thresholding cells at the QCSK receiver and the net concentration of the modulated signals received by the thresholding cells located along $\mathcal{S}_{i,j}$ is expressed as \begin{align}
    C_{\mathrm{Ch}}[t,t+t_{s}|\mathcal{S}_{i,j}] = \mathcal{P}\Big[C_{\mathrm{Tx}}^{S_1}(t,t+t_{s}),L_{3}-L_{2},\mathcal{S}_{3,4},\mathcal{S}_{i,j}\Big] + \mathcal{P}\Big[C_{\mathrm{Tx}}^{S_0}(t,t+t_{s}),L_{3}-L_{2},\mathcal{S}_{0,3},\mathcal{S}_{i,j}\Big].
\end{align} Based on \eqref{eq:T}, the net concentration of $\alpha$Sc molecules generated by thresholding unit $B_{j}$ can be derived as \vspace{-0.5cm}\begin{align}
    C_{B_{l}}(t,t+t_{s}|\mathcal{S}_{i,j}) = \mathcal{T}\Big[C_{\mathrm{Ch}}[t,t+t_{s}|\mathcal{S}_{i,j}], C_{\mathrm{Th}}^{B_l}(t), \mathrm{DOX}, \alpha\mathrm{Sc} \Big].
\end{align} The $\alpha$Sc molecules generated from the thresholding cells propagate to the back-end module of the receiver, which is responsible for processing the signal back into $Y_{1}$ and $Y_{0}$. From \eqref{eq:Y1boolean}, the output of bit $Y_{1}$ is the signal processing of double NOT operation on $B_{1}$ signal. Thus, the net concentration of $\alpha$Sc molecules for bit $Y_{1}$ detected at $x = L_{9}$ by the detection wall can be expressed as  \begin{align}
    C_{Y_{1}}(t,t+t_{s})=&\mathcal{P}\bigg\{\aunderbrace[l3D1@{~\ldots}]{\mathcal{N}\bigg[\mathcal{P}\Big\{ \underbrace{\mathcal{N}\Big[ \mathcal{P}[C_{B_{1}}(t,t+t_{s}|\mathcal{S}_{3,4}),L_{5}-L_{4},\mathcal{S}_{3,4},\mathcal{S}_{3,4}],\alpha\mathrm{Sc},\alpha\mathrm{Ca} \Big]}_{\text{First population of NOT cells: }\overline{B_{1}}}}_{\text{\scriptsize Second population of NOT cells: $\overline{\overline{B_{1}}}$}} \nonumber\\&\aunderbrace[@{~\ldots}1r]{, L_{7}-L_{6} ,\mathcal{S}_{3,4} ,\mathcal{S}_{3,4}\Big\} ,\alpha\mathrm{Ca} ,\alpha\mathrm{Sc}\bigg]}, L_{9}-L_{8}, \mathcal{S}_{3,4},\mathcal{S}_{3,4}\bigg\} .\label{eq:QCSKY1}
\end{align} For output bit $Y_{0}$, it is the mix of $\overline{\overline{B_{2}}}$ and $\overline{B_{1} + \overline{B_{0}}}$ as in \eqref{eq:Y0boolean}. Similar to \eqref{eq:QCSKY1}, the output of $\overline{\overline{B_{2}}}$ is generated through a double NOT operation at $x = L_{8}$, which can be expressed as  \begin{align}
    &C_{1}(t,t+t_{s}) =\nonumber\\ 
    &\underbrace{\mathcal{N}\bigg[\mathcal{P}\Big\{ \underbrace{\mathcal{N}\Big[\mathcal{P}[C_{B_{2}}(t,t+t_{s}|\mathcal{S}_{2,3}), L_{5}-L_{4}, \mathcal{S}_{2,3}, \mathcal{S}_{2,3}],\alpha\mathrm{Sc},\alpha\mathrm{Ca}\Big]}_{\text{First population of NOT cells: }\overline{B_{2}}},L_{7}-L_{6} , \mathcal{S}_{2,3}, \mathcal{S}_{2,3}\Big\},\alpha\mathrm{Ca},\alpha\mathrm{Sc}\bigg]}_{\text{Second population of NOT cells: }\overline{\overline{B_{2}}}}\label{eq:QCSKnnB2}.
\end{align} The output of $\overline{B_{1} + \overline{B_{0}}}$ at $x=L_{8}$ can be expressed as a NOT operation of a summation of the concentration from $B_{1}$ and $\overline{B_{0}}$, which is  \begin{align}
    &C_{2}(t,t+t_{s}) =\nonumber\\ 
    &\mathcal{N}\bigg[ 
    \mathcal{P}\Big\{ \underbrace{\mathcal{I}\Big[ \mathcal{P}[C_{B_{1}}(t,t+t_{s}|\mathcal{S}_{1,2}),L_{5}-L_{4},\mathcal{S}_{1,2},\mathcal{S}_{1,2}],\alpha\mathrm{Sc},\alpha\mathrm{Ca}\Big]}_{\text{First population of ID cells: } B_{1}} ,L_{7}-L_{6},\mathcal{S}_{1,2},\mathcal{S}_{0,2} \Big\}    
    \nonumber\\&+ \mathcal{P}\Big\{ \underbrace{\mathcal{N}\Big[\mathcal{P}[C_{B_{0}}(t,t+t_{s}|\mathcal{S}_{0,1}),L_{5}-L_{4},\mathcal{S}_{0,1},\mathcal{S}_{0,1}] ,\alpha\mathrm{Sc},\alpha\mathrm{Ca}\Big]}_{\text{First population of NOT cells: } \overline{B_{0}}},L_{7}-L_{6},\mathcal{S}_{0,1},\mathcal{S}_{0,2} \Big\}, \alpha\mathrm{Ca}, \alpha\mathrm{Sc}\bigg]. \label{eq:QCSKB1B0}
\end{align} Therefore, the output of $Y_{0}$ detected at $x=L_{9}$ can be expressed as a summation of $\alpha$Sc molecules propagated from $x=L_{8}$ as  \begin{align}
    &C_{Y_{0}}(t,t+t_{s}) = 
    \mathcal{P}[C_{1}(t,t+t_{s}),L_{9}-L_{8} , \mathcal{S}_{2,3},\mathcal{S}_{0,3} ] + \mathcal{P}[C_{2}(t,t+t_{s}), L_{9}-L_{8}, \mathcal{S}_{0,2},\mathcal{S}_{0,3} ]. \label{eq:QCSKY0}
\end{align}

\vspace{-0.5cm}
\section{Numerical Results}\label{sec:NumericalResults}
In this section, we simulate our proposed BCSK and QCSK systems in BSim and Matlab to examine the effectiveness of our CSK design framework and validate our theoretical analysis. The BSim is a computational agent-based simulator for modeling cell dynamics in a typical environment \cite{Matyjaszkiewicz2017BSimSimulator}. In BSim, the propagation of molecules is simulated using particle-based method \cite{Cai2006ModellingMigration, Deng2016MolecularReceiver}, and each engineered cell is considered as an agent and simulated using an ODE solver. By contrast, in MATLAB, all engineered cells in a population are considered as a single agent. The simulation time interval $t_{s}$ is set as 0.01s and the results are averaged over 2000 realizations. We set the molecules propagation parameters in \eqref{eq:ard2} as $D = 89 \mu \textrm{m}^{2}\; \textrm{s}^{-1}$, $u = 0.1 \mu \textrm{m}^{2}\; \textrm{s}^{-1}$, and $k_{\mathrm{a}} = 9 \mu \textrm{m}\; \textrm{s}^{-1}$. We also set the parameters for engineered cell processes as $\xi = 20 \textrm{s}^{-1}$ and $k_{\mathrm{f}} = 1\text{nM s}^{-1}$. It is noted that as the input molecules of an engineered cell are those absorbed on the cell membrane, we set the molecule exchange rate in \eqref{eq:CinODE} as $\eta = 1$. The other parameters are summarized in Tables~\ref{tb:cellpara} and \ref{tb:parameter}. In all the figures, we also use “Ana.” and “Sim.” to abbreviate “Analytical” and “Simulation”, respectively.
\begin{table}[!t]
    \centering
        {\small
            \caption{The parameters of the engineered cells.}
            \label{tb:cellpara}
            \vspace{-0.4cm}    
            \begin{tabular}{|c|c|c|c|c|c|}\hline
                \textbf{Molecule}&$\beta\text{ (nM min}^{-1}$)&$\theta\text{ (nM}^{-1}$)&$n$&$k_{\mathrm{d}}$ ($\text{min}^{-1}$)&Reference\\ \hline
                $\alpha$Ca&0.0369&0.26&0.9&0.05&\cite{Urrios2016ADevice}\\
                $\alpha$Sc&0.162&0.167&1.2&0.023&\cite{Urrios2016ADevice}\\
                DOX&0.162&0.167&1.2&0.023&-\\
                R (Repressor)&0.615&1550&2&0.15&\cite{Urrios2016ADevice}\\\hline
            \end{tabular}\\}
\end{table}
\begin{table}[!t]
    \centering {\small
    \caption{The dimension of the CSK systems.}
    \label{tb:parameter}
    \vspace{-0.4cm}
    \begin{tabular}{|c|c|c|c|c|c|c|c|c|c|c|c|c|c|c|c|c|}
        \hline \textbf{Parameter}&$L_0$&$L_1$&$L_2$&$L_3$&$L_4$&$L_5$&$L_6$&$L_7$&$L_8$&$L_9$&$W_{0}$&$W_1$&$W_2$&$W_3$&$W_4$&$H$\\ \hline
        \textbf{Value ($\mu\mathrm{m}$)}&0&1&4&39&42&43&46&47&50&55&0&2.5&5&10&15&3\\\hline 
    \end{tabular}}
\end{table}
\vspace{-0.5cm}
\subsection{BSim vs. MATLAB}\label{sec:efProp}
\begin{figure}
    \centering
    \subfloat[The schematic of the simulation. \label{fig:propdia}]{\includegraphics[width=0.365\linewidth]{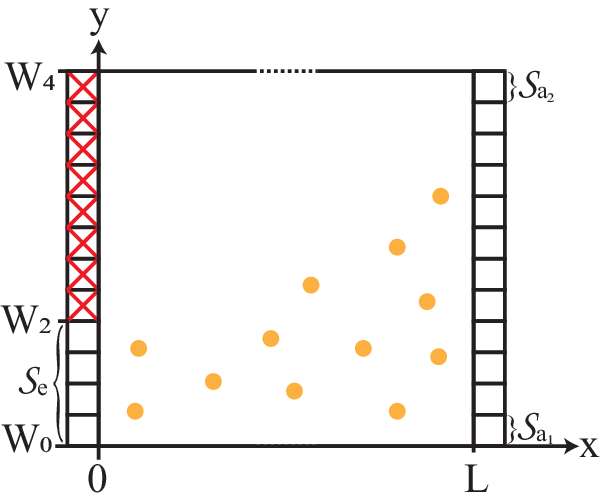}}\hspace{0.5cm}
    \subfloat[The reception of molecules at $L= 10$.\label{fig:propdiff}]{\includegraphics[scale = 0.4]{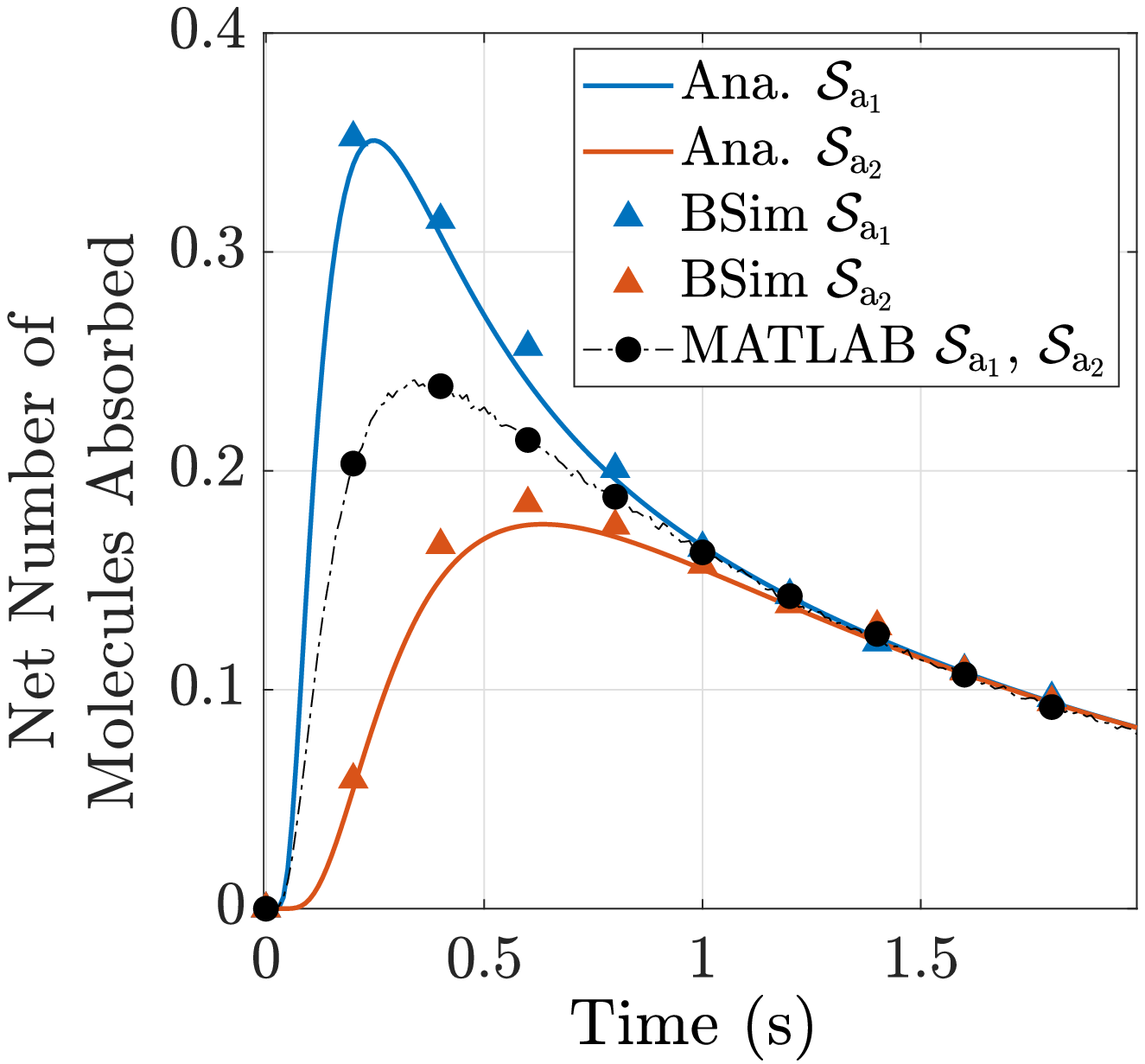}}
    \caption{The comparison of BSim and MATLAB simulators. The molecules are released by the cells located along $S_{\mathrm{e}}$ with ${W_{0} \leq y \leq W_{2}}$ and received by the cell at $x=L$ on $\mathcal{S}_{\mathrm{a}_{1}}$ and $\mathcal{S}_{\mathrm{a}_{2}}$ with ${0 \leq y \leq 1.25}$ and $13.75 \leq y \leq 15$, respectively. The cells at $x=0$ along $W_{2}\leq y \leq W_{4}$ (indicated as a box with a red cross) are not active and no molecules are released by those cells.}
    \label{fig:PropResult}
\end{figure}
We first investigate the simulation accuracy of BSim and Matlab. Specifically, as illustrated in \subfigref{fig:PropResult}{fig:propdia}, only the cells located along $S_{\mathrm{e}}$ at $x=0$ emit molecules into the channel, and two absorbing surfaces $\mathcal{S}_{\mathrm{a}_{1}}$ and $\mathcal{S}_{\mathrm{a}_{2}}$ are considered. The simulation results are averaged over 10,000 independent impulse emissions of 500 DOX molecules by the cells at time $t=0$. \subfigref{fig:PropResult}{fig:propdiff} plots the simulation results by BSim and MATLAB and the corresponding analytical values. We observed that the simulation results obtained from BSim have a close match to the analytical values whereas MATLAB simulations fail to capture the number of absorbed molecules. The reason for this difference is that the MATLAB simulates all the cells in the population located at $x=0$ or $x=L$ as a single agent, therefore the number of absorbed molecules for $\mathcal{S}_{\mathrm{a}_{1}}$ and $\mathcal{S}_{\mathrm{a}_{2}}$ are the same, which ignores that the molecules released by $S_{\mathrm{e}}$ need more time to arrive at $\mathcal{S}_{\mathrm{a}_{2}}$. This signifies the importance of an agent-based simulator to capture the individual behavior of each cell. Based on this, only BSim is used for the rest of the simulations. 
\vspace{-0.5cm}
\subsection{BCSK System}\label{sec:BCSKresult}
\begin{figure}
    \centering
    \subfloat[BCSK transmitter.\label{fig:BCSKTx}]{\includegraphics[scale = 0.4]{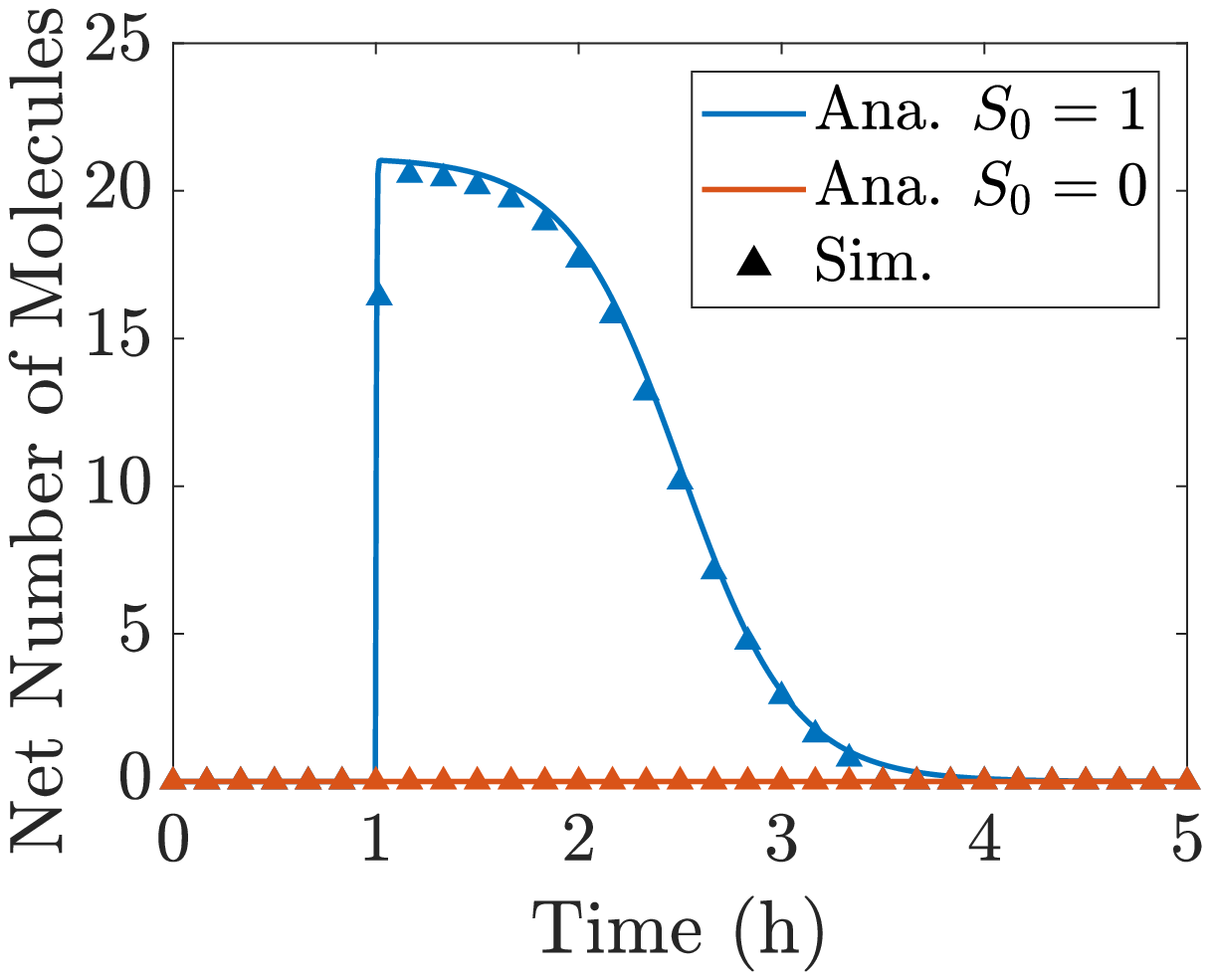}}\hspace{1cm}
    \subfloat[BCSK receiver.\label{fig:BCSKRx}]{\includegraphics[scale = 0.4]{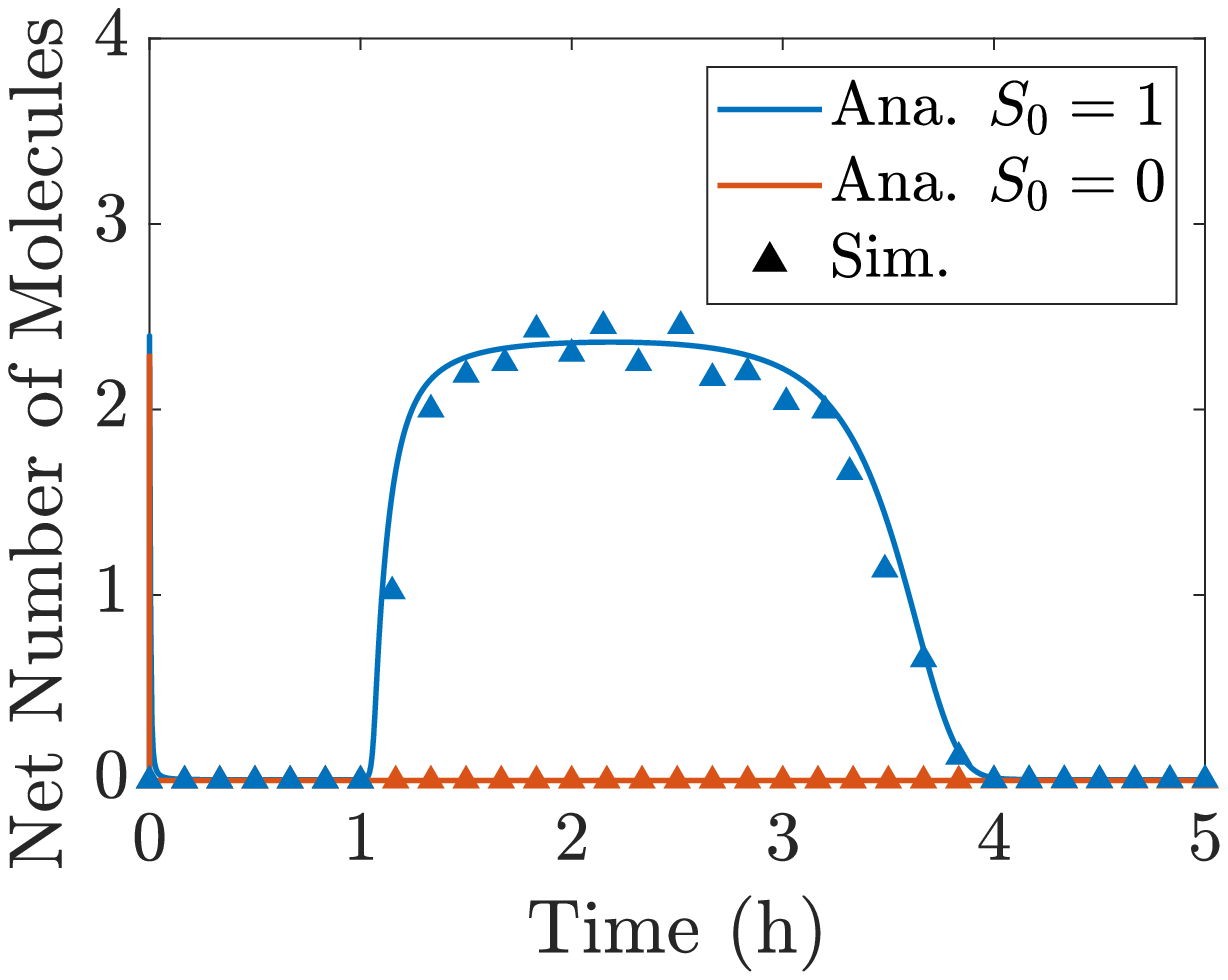}}
    \caption{The performance of our proposed BCSK system.}
    \label{fig:BCSKResult}
\end{figure}

To evaluate our proposed BCSK transceiver design in \figref{fig:BCSK}, a rectangular input with the amplitude of 750 $\alpha$Ca molecules (50nM) and a duration of 10s are released to the BCSK transmitter at $t=1$h to represent bit-1, i.e., $S_{0}=1$. We set the width of the channel as $W_{1} = 5\mu\text{m}$. For the thresholding unit, we set the concentration of threshold molecules as 0.01nM. \figref{fig:BCSKResult} plots the simulation results of our proposed BCSK system obtained from BSim and the analytical values obtained from \eqref{eq:BCSKTx} and \eqref{eq:BCSKRx}. We observe that there is a close match between the simulation points and the analytical curves, which validates our theoretical analysis. We also observe that our proposed transmitter releases a certain number of molecules after the addition of 50nM and no molecules are released; at the same time, it also shows that our proposed receiver is able to release information molecules in response to the signals from the OOK transmitter, demonstrating the effectiveness of OOK modulation and demodulation functionalities.
\vspace{-0.5cm}
\subsection{QCSK System}\label{sec:QCSKresult}
\enlargethispage{\baselineskip}
\vspace{-0.5cm}
\begin{figure}[!h]
    \centering
    \includegraphics[scale = 0.4]{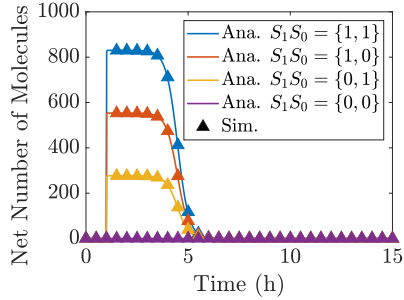}
    \caption{The performance of our proposed QCSK transmitter.}
    \label{fig:QCSKTx}
\end{figure}
\vspace{-1cm}To examine the performance of the QCSK design in \figref{fig:QCSK}, we consider all possible input combinations: $S_{1}S_{0} = \{1,1\}$, $S_{1}S_{0} = \{1,0\}$, $S_{1}S_{0} = \{0,1\}$, and $S_{1}S_{0} = \{0,0\}$. For bit-1 input, rectangular inputs with amplitude of 50nM\footnote{With respect to the geometry provided in \tabref{tb:parameter}, the amplitude is equivalent to 750 $\alpha$Ca molecules and 1500 $\alpha$Ca molecules for input bits $S_{0}$ and $S_{1}$, respectively.} of $\alpha$Ca molecules are released to the QCSK transmitter at $t=1$h. As the information molecules degrade during the propagation and cell signal processing, we increase the duration of the rectangular inputs to 30 mins. We also increase all the production rates by 40 folds, which can be achieved by amplifying circuits in practice \cite{mogas20212d}. For the thresholding unit, we set the thresholding parameters as $C_{\mathrm{Th}}^{B_2}(t) = 0.7$nM, $C_{\mathrm{Th}}^{B_1}(t) = 0.45$nM, and $C_{\mathrm{Th}}^{B_0}(t) = 0.1$nM. \figref{fig:QCSKTx} plots the simulation results of our proposed QCSK transmitter and the analytical results calculated by \eqref{eq:QCSKTx}. We observe that the QCSK transmitter can modulate two bits into four different concentration levels, thus achieving QCSK modulation. \figref{fig:QCSKResult} plots the simulation results of our proposed QCSK receiver and analytical results in \eqref{eq:QCSKY1} and \eqref{eq:QCSKY0}. It is clear that the receiver is capable of distinguishing the concentration levels and demodulating the signals back to two bits. Furthermore, both figures present a close match between the analytical and simulation results, which further validates our theoretical analysis. 

\begin{figure}
    \centering
    \subfloat[$S_1S_0 = \{1,1\}$\label{fig:QCSK11}]{\includegraphics[scale = 0.3]{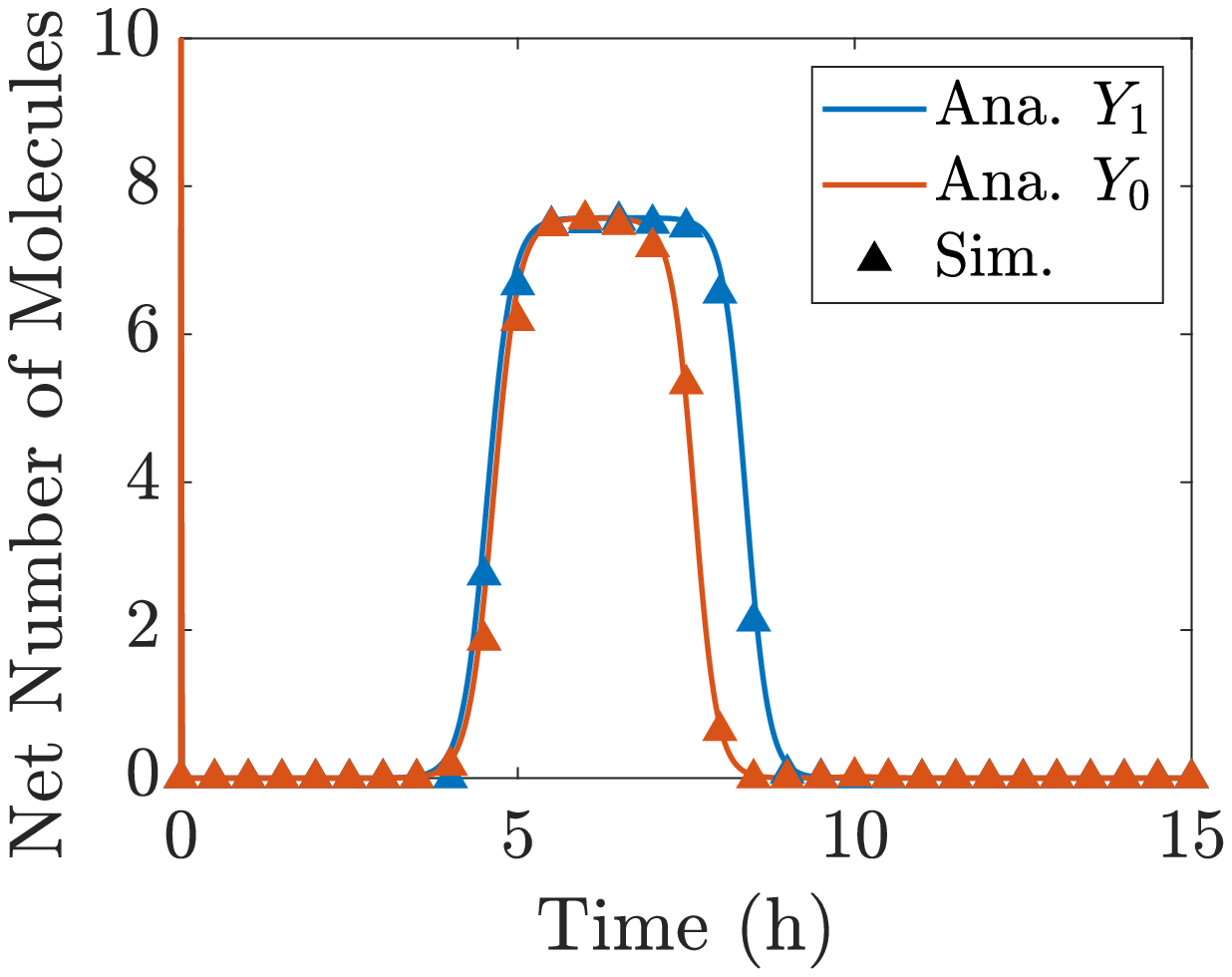}}\hspace{0.1cm}
    \subfloat[$S_1S_0 = \{1,0\}$\label{fig:QCSK10}]{\includegraphics[scale = 0.3]{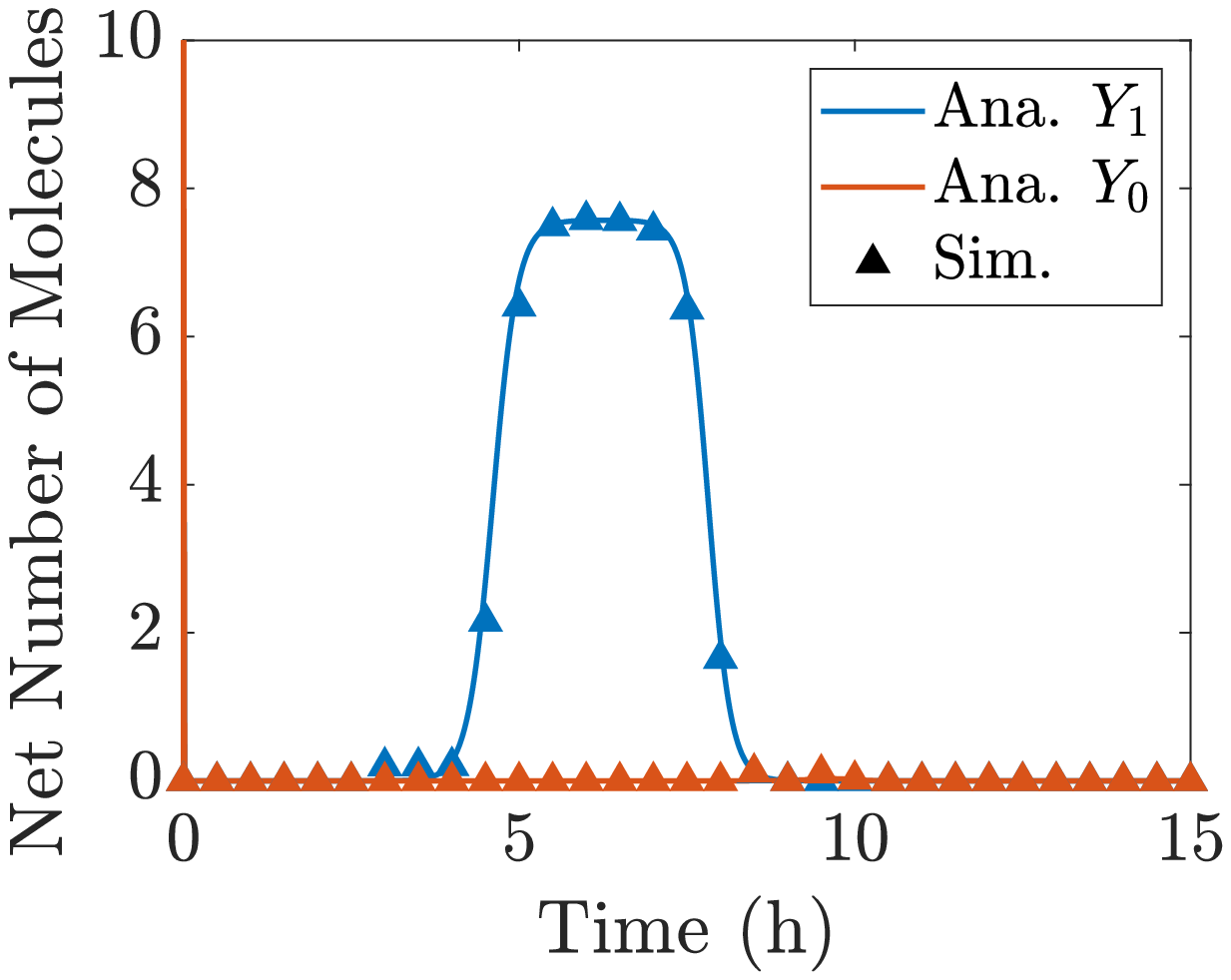}}\hspace{0.1cm}
    \subfloat[$S_1S_0 = \{0,1\}$\label{fig:QCSK01}]{\includegraphics[scale = 0.3]{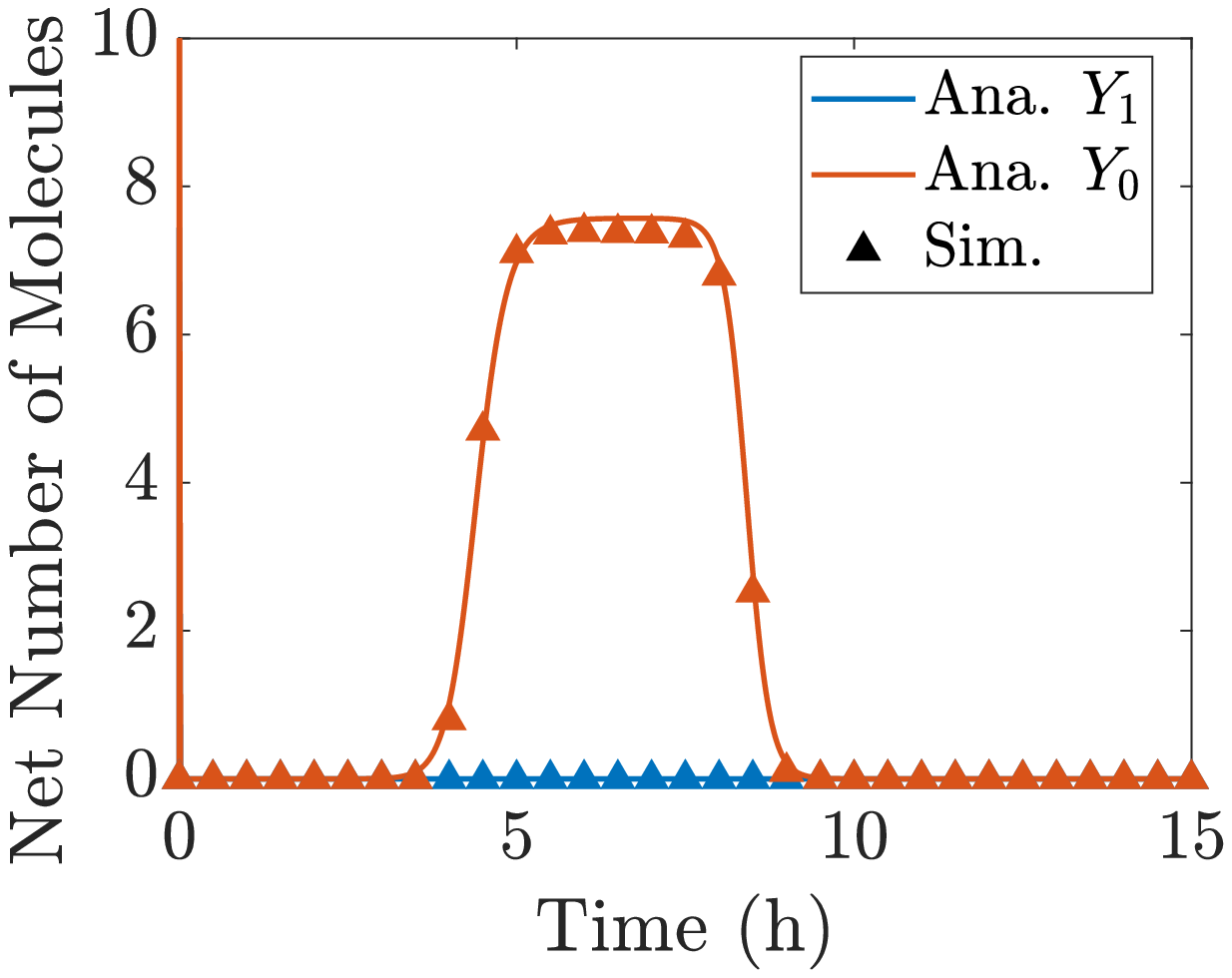}}\hspace{0.1cm}
    \subfloat[$S_1S_0 = \{0,0\}$\label{fig:QCSK00}]{\includegraphics[scale = 0.3]{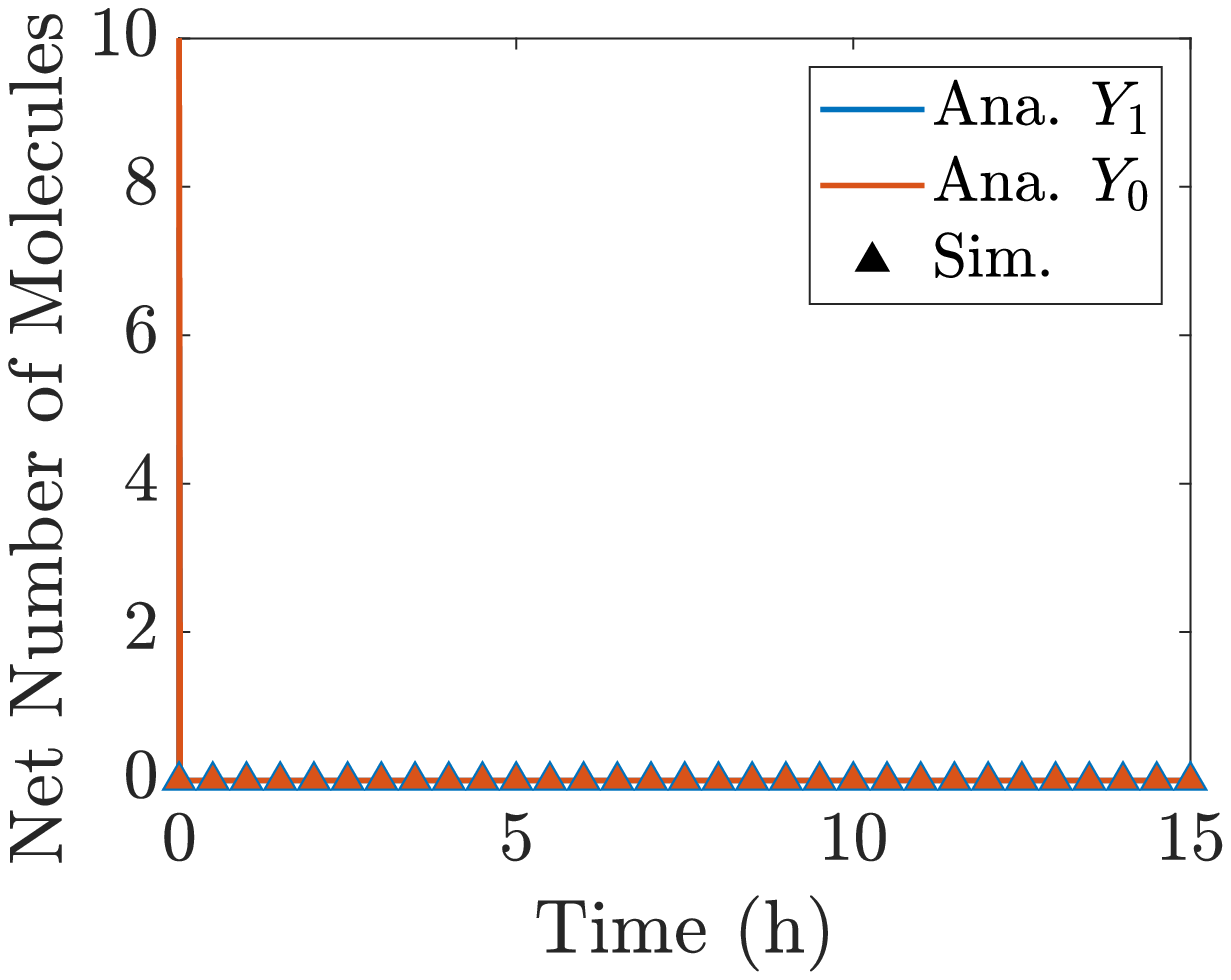}}
    \caption{The performance of our proposed QCSK receiver.}
    \label{fig:QCSKResult}
\end{figure}

\vspace{-0.5cm}
\subsection{Communication Performance of QCSK Design}\label{sec:ComPerf}
\vspace{-0.5cm}\begin{figure}[!h]
    \centering
    \includegraphics[scale = 0.35]{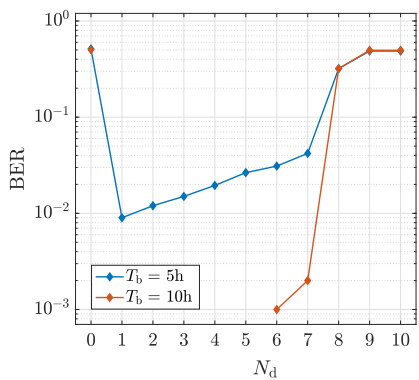}
    \caption{Investigation of the impact of detection parameter on the QCSK performance.}
    \label{fig:QCSKPerformance}
\end{figure}
\vspace{-1cm}We also examine the communication performance of our proposed QCSK system. In particular, a 1000-bit sequence is randomly generated and transmitted under different bit intervals $T_{\mathrm{b}}$ and the simulation parameters for the QCSK system in Sec. IV-C are also used. According to the QCSK receiver performance in \figref{fig:QCSKResult}, the average response of bit-1 is roughly 8 molecules and appears at 5h after the start of the transmission; therefore, we sample the output bits 5h after each transmission. \figref{fig:QCSKPerformance} plots the BER against the detection thresholding value $N_{\mathrm{d}}$, which is the number of absorbed molecules required for the receiver to decide whether bit-1 is transmitted. For $T_{\mathrm{b}} = 5$h, we first observe that when $N_{\mathrm{d}} = 0$, the maximum BER for QCSK is obtained because all bits are detected as bit-1. The BER increases as $N_{\mathrm{d}}$ approaches the average number of detected molecules for bit-1. When $N_{\mathrm{d}}$ is greater than the average detected molecules, all bits are detected as bit-0 and this results in the maximum BER. Moreover, we find that $S_{0}=1$ is often detected as bit-0 when $S_{1}S_{0} = \{0,1\}$ is transmitted after $S_{1}S_{0} = \{1,1\}$. The thresholding output $B_{1}$ for $Y_{0}$ is bit-1 when $S_{1}S_{0} = \{1,1\}$ while is bit-0 when $S_{1}S_{0} = \{0,1\}$ (see \figref{fig:QCSK}). As $S_{1}S_{0} = \{1,1\}$ leads to a higher concentration level than $S_{1}S_{0} = \{0,1\}$ and the higher received concentration the longer duration of the thresholding output, the bit-1 of $B_1$ when $S_{1}S_{0} = \{1,1\}$ can accumulate in the channel and interfere the bit-0 of $B_1$ when $S_{1}S_{0} = \{0,1\}$, thus causing the wrong detection of $S_0$. This error can be mitigated by increasing $T_{\mathrm{b}}$ to ensure all molecules from the previous bit undergo natural degradation before arriving $x=L_{7}$. From \figref{fig:QCSKPerformance}, we can observe that when $T_{\mathrm{b}} =10$h, the BER significantly decreases and no error is detected for $N_{\mathrm{d}} < 6$.\footnote{The BER for $N_{\mathrm{d}} < 6$ is presumably lower than $10^{-3}$, thus a 1000-bit sequence is insufficient to detect the error accurately.}

\vspace{-0.7cm}
\section{Conclusion}\label{sec:Conclusion}
In this paper, we proposed a design principle to realize the concentration shift keying (CSK) system based on multicellular consortia and spatial segregation. According to our design principle, we developed a small library of engineered cells to realize binary CSK (BCSK) and quadruple CSK (QCSK) systems. Notably, the gene expressions and the involved molecules in our proposed system were specified in the synthetic biology domain. Then, we established an analytical framework that divides a system into four building blocks to theoretically characterized the proposed systems. We derived the impulse response of each building block, and the cascade of those impulse responses resulted in the analysis of the proposed system. Simulation results from BSim showed that our systems responded appropriately to input signals and closely matched our theoretical analysis. Additionally, we demonstrated that intersymbol interference (ISI) can be mitigated through a careful selection of bit interval and detection thresholding value. As our CSK design principle is based on simple and reusable single-input single-output cells with spatial segregation and relies on digital electronics theory, this paper presents a novel methodology to design and analyze diverse and complex signal processing functions for molecular communication (MC) over the molecular domain. The future work can include the wet-lab experimental validation and the incorporation of parabolic flow and stochastic analysis for the mathematical framework.

\vspace{-1cm}\appendices
\allowdisplaybreaks
\section{Proof of \thref{th:CR4}}\label{ap:proofBB4}
    The dynamics of the repressor concentration in \eqref{eq:dCR4} can be divided into three processes using the operator splitting method \cite{Dadi2022QCSK} as 
     \begin{subequations}
        \begin{alignat}{3}
            &\text{Production:}& \dfrac{\mathrm{d}C_{\mathrm{R}}(t)}{\mathrm{d}t}&=f_{\mathrm{R}}(t).\label{eq:CRprod}\\
            &\text{Degradation:}&\dfrac{\mathrm{d}C_{\mathrm{R}}(t)}{\mathrm{d}t}&=- k_{\mathrm{d}}C_{\mathrm{R}}(t).\label{eq:CRdeg}\\
            &\text{Bimolecular Reaction:}\;\;\;\;\;\;\;\;\;\;\;&\dfrac{\mathrm{d}C_{\mathrm{R}}(t)}{\mathrm{d}t}&=- k_{\mathrm{f}}C_{\mathrm{Iin}}(t)C_{\mathrm{R}}(t).\label{eq:CRre}
        \end{alignat}
    \end{subequations}
    We assume that the processes occur in the order of production, degradation, and bimolecular reaction, where the initial concentration of each process is the molecule concentration that resulted from previous processes. For the time interval $[(m-1)t_{s}, mt_{s}]$, the initial concentration for the degradation is a superposition of molecules produced by the gene expression in \eqref{eq:CRprod} and the residual repressor from the bimolecular reaction in the previous time interval $[(m-2)t_{s}, (m-1)t_{s}]$, which can be expressed as \vspace{-0.3cm}\begin{align}
        t_{s}f_{\mathrm{R}}[m-1] + C_{\mathrm{R}}^{\mathrm{Th}}[m-1],\label{eq:Cr0}
    \end{align}
    where $f_{\mathrm{R}}[m-1] = f_{\mathrm{R}}\big((m-1)t_{s}\big)$. This portion of repressor experiences the natural degradation in \eqref{eq:CRdeg}, resulting in the initial concentration for the bimolecular reaction, which can be expressed as \vspace{-0.7cm} \begin{align}
        C_{\mathrm{R_0}} = \{t_{s}f_{\mathrm{R}}[m-1] + C_{\mathrm{R}}^{\mathrm{Th}}[m-1]\}e^{-k_{\mathrm{d}}t_{s}}.\label{eq:Cr02}
    \end{align}\vspace{-0.2cm} Then, to find the repressor concentration at time $mt_{s}$, three cases of initial concentration need to be considered: \begin{itemize}
        \item \textbf{Case 1:} When $C_{\mathrm{R_0}} = 0$ or $C_{\mathrm{I_0}} = 0$, there is no chemical reaction. Therefore, the concentration of the repressor molecules is equal to its initial concentration, i.e., $C_{\mathrm{R}}^{\mathrm{Th}}[m] = C_{\mathrm{R_0}}$.
        \item \textbf{Case 2:} When $C_{\mathrm{R_0}} \neq C_{\mathrm{I_0}}$, chemical reaction in \eqref{eq:chemreact} occurs. Due to the one-to-one stoichiometric relation between reactants, eq.~\eqref{eq:CRre} can be rewritten as\begin{align}
        \dfrac{\mathrm{d}C_{\mathrm{R}}^{\mathrm{Th}}(t)}{\mathrm{d}t} = - k_{\mathrm{f}}C_{\mathrm{R}}^{\mathrm{Th}}(t)\{C_{\mathrm{R}}^{\mathrm{Th}}(t) - (C_{\mathrm{R_0}} - C_{\mathrm{I_0}})\}. \label{eq:case2chemrate}
    \end{align}
    By rearranging \eqref{eq:case2chemrate} and taking the integral during the time interval $[(m-1)t_{s},mt_{s}]$, the concentration of the repressor at $mt_{s}$ is derived as
    \begin{align}
        C_{\mathrm{R}}^{\mathrm{Th}}[m] = \dfrac{C_{\mathrm{R_0}}(C_{\mathrm{R_0}} - C_{\mathrm{I_0}})}{C_{\mathrm{R_0}}-C_{\mathrm{I_0}}\exp[-k_{\mathrm{f}}t_{s}(C_{\mathrm{R_0}}-C_{\mathrm{I_0}})]}.
    \end{align}
    \item \textbf{Case 3:} When $C_{\mathrm{R_0}} = C_{\mathrm{I_0}}$, eq.~\eqref{eq:case2chemrate} can be rewritten as\begin{align}
        \dfrac{\mathrm{d}C_{\mathrm{R}}^{\mathrm{Th}}(t)}{\mathrm{d}t} = - k_{\mathrm{f}}C^{2}_{\mathrm{R}}(t). \label{eq:case3rate}
    \end{align}
    By applying the same step as in \textbf{Case 2} on \eqref{eq:case3rate}, we arrive at \begin{align}
       C_{\mathrm{R}}^{\mathrm{Th}}[m] = \dfrac{C_{\mathrm{R_0}}}{1+k_{\mathrm{f}}t_{s}C_{\mathrm{R_0}}}.
    \end{align}
    \end{itemize} Combining those three cases with \eqref{eq:Cr02} as the initial condition, we arrive at \eqref{eq:CR4}.
\vspace{-0.7cm}\section{Proof of \thref{th:prop}}\label{ap:proofprop}
    We use the separation variable method \cite{jad2022BCSK} by assuming $C(\mathbf{d},t) = X(x,t)Y(y)Z(z)$ to find the impulse response of \eqref{eq:ard2}. The impulse response $h_{2}(\mathbf{d},t)$ of propagation block is derived as
    \begin{align}
        \hspace{-.4cm}h_{2}(\mathbf{d},t) =& \dfrac{2}{W}\sum\limits_{l=1}^{\infty} \dfrac{(\lambda^{2}_{l} + G_{1}^{2})\cos(\lambda_{l}x)}{L(\lambda^{2}_{l} + G_{1}^{2})+G_{1}}e^{\frac{u(2x-ut)}{4D}-\lambda_{l}^{2}Dt-k_\mathrm{d}t}\bigg[ 1 + 2\sum\limits_{i=1}^{\infty} \cos(\gamma_{i}y_{0})\cos(\gamma_{i}y)e^{-\gamma_{i}^{2}Dt}\bigg] \label{eq:h1}.
    \end{align} The impulse response of the molecule propagation to the absorbing boundary $\mathcal{S}_{\mathrm{a}}$ when the molecules are released by the emitting boundary $\mathcal{S}_{\mathrm{e}}$ can be expressed as \begin{align}
        h_{2}(t|L, \mathcal{S}_{\mathrm{e}}, \mathcal{S}_{\mathrm{a}}) =  \int_{y_{0_{1}}}^{y_{0_{2}}}\int_{0}^{H}\int_{y_{1}}^{y_{2}} h_{2}(\mathbf{d} = [L,y,z],\tau)  \mathrm{d}y\mathrm{d}z\mathrm{d}y_{0}\label{eq:h2t}.
    \end{align} The above equation can lead to \eqref{eq:h1boundary}. Thus, the concentration of the molecules can be expressed as \vspace{-0.7cm} \begin{align}
        C(t|L,\mathcal{S}_{\mathrm{e}},\mathcal{S}_{\mathrm{a}}) = C_{\mathrm{I}}(t) *h_{2}(t| L,\mathcal{S}_{\mathrm{e}}, \mathcal{S}_{\mathrm{a}}) \label{eq:L1Conc}.
    \end{align}
    To characterize the number of molecules absorbed on the absorbing wall, we define the change in the total number of particles due to the particle flux as \eqcite{Schulten2000LecturesBiophysics}{3.20} \begin{align}
        \dfrac{\mathrm{d}N(t)}{\mathrm{d}t} = \int_{\mathcal{S}} D\nabla C(\mathbf{d},t)\mathrm{d}\mathcal{S} \label{eq:ChangeinNum},
    \end{align} where $\mathcal{S}$ is the surface of the given region with differential element $\mathrm{d}\mathcal{S}$. The change in the total number of particles in \eqref{eq:ChangeinNum} on the absorbing wall is equivalent to the flux at the boundary specified in \eqref{eq:AbWallBC}; therefore, we can arrive at  \begin{align}
        N(t|L,\mathcal{S}_{\mathrm{e}}, \mathcal{S}_{\mathrm{a}}) = \int_{0}^{t} k_{\mathrm{a}}C(\tau|L,\mathcal{S}_{\mathrm{e}},\mathcal{S}_{\mathrm{a}})\mathrm{d}\tau\label{eq:Nreact}.
    \end{align} Dividing \eqref{eq:Nreact} by the cross-section of the detection area $H(y_{2}-y_{1})$ to translate from the number of molecules to concentration and taking the net change, we can arrive at \eqref{eq:gateOut}.

\singlespacing


\begin{thebibliography}{10}
\providecommand{\url}[1]{#1}
\csname url@samestyle\endcsname
\providecommand{\newblock}{\relax}
\providecommand{\bibinfo}[2]{#2}
\providecommand{\BIBentrySTDinterwordspacing}{\spaceskip=0pt\relax}
\providecommand{\BIBentryALTinterwordstretchfactor}{4}
\providecommand{\BIBentryALTinterwordspacing}{\spaceskip=\fontdimen2\font plus
\BIBentryALTinterwordstretchfactor\fontdimen3\font minus
  \fontdimen4\font\relax}
\providecommand{\BIBforeignlanguage}[2]{{%
\expandafter\ifx\csname l@#1\endcsname\relax
\typeout{** WARNING: IEEEtran.bst: No hyphenation pattern has been}%
\typeout{** loaded for the language `#1'. Using the pattern for}%
\typeout{** the default language instead.}%
\else
\language=\csname l@#1\endcsname
\fi
#2}}
\providecommand{\BIBdecl}{\relax}
\BIBdecl

\bibitem{jad2022BCSK}
R.~Jadsadaphongphaibool, D.~Bi, and Y.~Deng, ``{Binary Concentration Shift
  Keying Realization for Molecular Communication via Genetic Circuits},'' in
  \emph{Proc. ACM NanoCom}, Oct. 2022, pp. 1--6.

\bibitem{farsad2016comprehensive}
N.~Farsad, H.~B. Yilmaz, A.~Eckford, C.-B. Chae, and W.~Guo, ``A comprehensive
  survey of recent advancements in molecular communication,'' \emph{IEEE
  Commun. Surv. Tutor.}, vol.~18, no.~3, pp. 1887--1919, Feb. 2016.

\bibitem{pierobon2010physical}
M.~Pierobon and I.~F. Akyildiz, ``A physical end-to-end model for molecular
  communication in nanonetworks,'' \emph{IEEE J. Sel. Areas Commun.}, vol.~28,
  no.~4, pp. 602--611, May 2010.

\bibitem{Deng2016MolecularReceiver}
Y.~Deng, A.~Noel, M.~Elkashlan, A.~Nallanathan, and K.~C. Cheung, ``Modeling
  and simulation of molecular communication systems with a reversible
  adsorption receiver,'' \emph{IEEE Trans. Mol. Biol. Multi-Scale Commun.},
  vol.~1, no.~4, pp. 347--362, Dec. 2015.

\bibitem{kadloor2012molecular}
S.~Kadloor, R.~S. Adve, and A.~W. Eckford, ``Molecular communication using
  brownian motion with drift,'' \emph{IEEE Trans. Nanobiosci.}, vol.~11, no.~2,
  pp. 89--99, Jun. 2012.

\bibitem{nakano2012channel}
T.~Nakano, Y.~Okaie, and J.-Q. Liu, ``Channel model and capacity analysis of
  molecular communication with brownian motion,'' \emph{IEEE Commun. Lett.},
  vol.~16, no.~6, pp. 797--800, Jun. 2012.

\bibitem{kilinc2013receiver}
D.~Kilinc and O.~B. Akan, ``Receiver design for molecular communication,''
  \emph{IEEE J. Sel. Areas Commun.}, vol.~31, no.~12, pp. 705--714, Dec. 2013.

\bibitem{kuscu2019transmitter}
M.~Kuscu, E.~Dinc, B.~A. Bilgin, H.~Ramezani, and O.~B. Akan, ``Transmitter and
  receiver architectures for molecular communications: A survey on physical
  design with modulation, coding, and detection techniques,'' \emph{Proc.
  IEEE}, vol. 107, no.~7, pp. 1302--1341, Jul. 2019.

\bibitem{soldner2020survey}
C.~A. S{\"o}ldner, E.~Socher, V.~Jamali, W.~Wicke, A.~Ahmadzadeh, H.-G.
  Breitinger, A.~Burkovski, K.~Castiglione, R.~Schober, and H.~Sticht, ``A
  survey of biological building blocks for synthetic molecular communication
  systems,'' \emph{IEEE Commun. Surv. Tutor.}, vol.~22, no.~4, pp. 2765--2800,
  Jul. 2020.

\bibitem{bi2021survey}
D.~Bi, A.~Almpanis, A.~Noel, Y.~Deng, and R.~Schober, ``{A survey of molecular
  communication in cell biology: Establishing a new hierarchy for
  interdisciplinary applications},'' \emph{IEEE Commun. Surv. Tutor.}, vol.~23,
  no.~3, pp. 1494--1545, Mar. 2021.

\bibitem{Kuran2021ADiffusion}
M.~S. Kuran, H.~B. Yilmaz, I.~Demirkol, N.~Farsad, and A.~Goldsmith, ``{A
  Survey on Modulation Techniques in Molecular Communication via Diffusion},''
  \emph{IEEE Commun. Surv. Tutor.}, vol.~23, no.~1, pp. 7--28, Jan. 2021.

\bibitem{Dadi2022QCSK}
D.~Bi and Y.~Deng, ``{Microfluidic QCSK Transmitter and Receiver Design for
  Molecular Communication},'' \emph{IEEE Trans. Commun.}, vol.~70, no.~9, pp.
  5837--5852, Sep. 2022.

\bibitem{bi2020chemical}
D.~Bi, Y.~Deng, M.~Pierobon, and A.~Nallanathan, ``Chemical reactions-based
  microfluidic transmitter and receiver design for molecular communication,''
  \emph{IEEE Trans. Commun.}, vol.~68, no.~9, pp. 5590--5605, Sep. 2020.

\bibitem{bi2021digital}
D.~Bi and Y.~Deng, ``Digital signal processing for molecular communication via
  chemical-reaction-based microfluidic circuits,'' \emph{IEEE Commun. Mag.},
  vol.~59, no.~5, pp. 26--32, May 2021.

\bibitem{khalil2010synthetic}
A.~S. Khalil and J.~J. Collins, ``Synthetic biology: applications come of
  age,'' \emph{Nat. Rev. Genet.}, vol.~11, no.~5, pp. 367--379, May 2010.

\bibitem{tamsir2011robust}
A.~Tamsir, J.~J. Tabor, and C.~A. Voigt, ``{Robust multicellular computing
  using genetically encoded NOR gates and chemical ‘wires’},''
  \emph{Nature}, vol. 469, no. 7329, pp. 212--215, Dec. 2011.

\bibitem{wang2011engineering}
B.~Wang, R.~I. Kitney, N.~Joly, and M.~Buck, ``{Engineering modular and
  orthogonal genetic logic gates for robust digital-like synthetic biology},''
  \emph{Nat. Commun.}, vol.~2, no.~1, pp. 1--9, Oct. 2011.

\bibitem{gardner2000construction}
T.~S. Gardner, C.~R. Cantor, and J.~J. Collins, ``{Construction of a genetic
  toggle switch in Escherichia coli},'' \emph{Nature}, vol. 403, no. 6767, pp.
  339--342, Jan. 2000.

\bibitem{Urrios2016ADevice}
A.~Urrios, J.~Mac{\'\i}a, R.~Manzoni, N.~Conde, A.~Bonforti, E.~de~Nadal,
  F.~Posas, and R.~Sol{\'{e}}, ``{A Synthetic Multicellular Memory Device},''
  \emph{ACS Synth. Biol.}, vol.~5, no.~8, pp. 862--873, Aug. 2016.

\bibitem{unluturk2015biotranceiver}
B.~D. Unluturk, A.~O. Bicen, and I.~F. Akyildiz, ``{Genetically Engineered
  Bacteria-Based BioTransceivers for Molecular Communication},'' \emph{IEEE
  Trans. Commun.}, vol.~63, no.~4, pp. 1271--1281, Feb. 2015.

\bibitem{Marcone2018Parity-checkCells}
A.~Marcone, M.~Pierobon, and M.~Magarini, ``{Parity-check coding based on
  genetic circuits for engineered molecular communication between biological
  cells},'' \emph{IEEE Trans. Commun.}, vol.~66, no.~12, pp. 6221--6236, Dec.
  2018.

\bibitem{nielsen2016genetic}
A.~A. Nielsen, B.~S. Der, J.~Shin, P.~Vaidyanathan, V.~Paralanov, E.~A.
  Strychalski, D.~Ross, D.~Densmore, and C.~A. Voigt, ``{Genetic circuit design
  automation},'' \emph{Sci.}, vol. 352, no. 6281, p. aac7341, Apr. 2016.

\bibitem{gander2017digital}
M.~W. Gander, J.~D. Vrana, W.~E. Voje, J.~M. Carothers, and E.~Klavins,
  ``Digital logic circuits in yeast with crispr-dcas9 nor gates,'' \emph{Nat.
  Commun.}, vol.~8, no.~1, pp. 1--11, May 2017.

\bibitem{macia2012distributed}
J.~Mac{\'\i}a, F.~Posas, and R.~V. Sol{\'e}, ``Distributed computation: the new
  wave of synthetic biology devices,'' \emph{Trends Biotechnol.}, vol.~30,
  no.~6, pp. 342--349, Jun. 2012.

\bibitem{Al-Radhawi2020DistributedCircuits}
M.~A. Al-Radhawi, A.~P. Tran, E.~A. Ernst, T.~Chen, C.~A. Voigt, and E.~D.
  Sontag, ``{Distributed Implementation of Boolean Functions by Transcriptional
  Synthetic Circuits},'' \emph{ACS Synth. Biol.}, vol.~9, no.~8, pp.
  2172--2187, Aug. 2020.

\bibitem{Macia2016ImplementationConsortia}
J.~Mac{\'\i}a, R.~Manzoni, N.~Conde, A.~Urrios, E.~de~Nadal, R.~Sol{\'{e}}, and
  F.~Posas, ``{Implementation of Complex Biological Logic Circuits Using
  Spatially Distributed Multicellular Consortia},'' \emph{PLoS Comput. Biol.},
  vol.~12, no.~2, p. e1004685, Feb. 2016.

\bibitem{kuran2011modulation}
M.~S. Kuran, H.~B. Yilmaz, T.~Tugcu, and I.~F. Akyildiz, ``Modulation
  techniques for communication via diffusion in nanonetworks,'' in \emph{Proc.
  IEEE ICC}, Jun. 2011, pp. 1--5.

\bibitem{kettani2006conversion}
H.~Kettani, ``On the conversion between number systems,'' \emph{IEEE Trans.
  Circuits Syst. II}, vol.~53, no.~11, pp. 1255--1258, Nov. 2006.

\bibitem{Urrios2018Plug-and-PlayResponses}
A.~Urrios, E.~Gonzalez-Flo, D.~Canadell, E.~De~Nadal, J.~Mac{\'\i}a, and
  F.~Posas, ``{Plug-and-Play Multicellular Circuits with Time-Dependent Dynamic
  Responses},'' \emph{ACS Synth. Biol.}, vol.~7, no.~4, pp. 1095--1104, Apr.
  2018.

\bibitem{ciletti2007digital}
M.~D. Ciletti and M.~M. Mano, \emph{Digital design: with an introduction to the
  Verilog HDL}.\hskip 1em plus 0.5em minus 0.4em\relax New Jersey, USA:
  Prentice-Hall, 2007.

\bibitem{fiore2017silico}
G.~Fiore, A.~Matyjaszkiewicz, F.~Annunziata, C.~Grierson, N.~J. Savery,
  L.~Marucci, and M.~Di~Bernardo, ``In-silico analysis and implementation of a
  multicellular feedback control strategy in a synthetic bacterial
  consortium,'' \emph{ACS Synth. Biol.}, vol.~6, no.~3, pp. 507--517, Dec.
  2017.

\bibitem{Ingalls2013MathematicalIntroduction}
B.~P. Ingalls, \emph{{Mathematical modeling in systems biology: an
  introduction}}.\hskip 1em plus 0.5em minus 0.4em\relax Cambridge,
  Massachusetts, USA: MIT press, Jul. 2013.

\bibitem{Jamali2019Channel}
V.~Jamali, A.~Ahmadzadeh, W.~Wicke, A.~Noel, and R.~Schober, ``{Channel
  Modeling for Diffusive Molecular Communication—A Tutorial Review},''
  \emph{Proc. IEEE}, vol. 107, no.~7, pp. 1256--1301, Jul. 2019.

\bibitem{kumar2010microfluidic}
C.~S. Kumar, \emph{{Microfluidic devices in nanotechnology: Fundamental
  concepts}}.\hskip 1em plus 0.5em minus 0.4em\relax Hoboken, New Jersey, USA:
  John Wiley \& Sons, 2010.

\bibitem{Mosayebi2019EarlyNanosensors}
R.~Mosayebi, A.~Ahmadzadeh, W.~Wicke, V.~Jamali, R.~Schober, and
  M.~Nasiri-Kenari, ``{Early cancer detection in blood vessels using mobile
  nanosensors},'' \emph{IEEE Trans. Nanobiosci.}, vol.~18, no.~2, pp. 103--116,
  Apr. 2019.

\bibitem{Matyjaszkiewicz2017BSimSimulator}
A.~Matyjaszkiewicz, G.~Fiore, F.~Annunziata, C.~S. Grierson, N.~J. Savery,
  L.~Marucci, and M.~Di~Bernardo, ``{BSim 2.0: An Advanced Agent-Based Cell
  Simulator},'' \emph{ACS Synth. Biol.}, vol.~6, no.~10, pp. 1969--1972, Oct.
  2017.

\bibitem{Cai2006ModellingMigration}
A.~Q. Cai, K.~A. Landman, and B.~D. Hughes, ``{Modelling Directional Guidance
  and Motility Regulation in Cell Migration},'' \emph{Bull. Math. Biol. 2005
  68:1}, vol.~68, no.~1, pp. 25--52, Mar. 2006.

\bibitem{mogas20212d}
S.~Mogas-D{\'\i}ez, E.~Gonzalez-Flo, and J.~Mac{\'\i}a, ``{2D printed
  multicellular devices performing digital and analogue computation},''
  \emph{Nat. Commun.}, vol.~12, no.~1, pp. 1--10, Mar. 2021.

\bibitem{Schulten2000LecturesBiophysics}
K.~Schulten and I.~Kosztin, \emph{{Lectures in Theoretical Biophysics}}.\hskip
  1em plus 0.5em minus 0.4em\relax Urbana, Illinois, USA: University of
  Illinois, Apr. 2000.

\end{thebibliography}
\end{document}